\documentclass[twocolumn]{aastex631} 

\usepackage{amsmath}
\usepackage{booktabs}
\usepackage{xcolor}
\usepackage{graphicx}
\usepackage{tabularx}
\usepackage{hyperref}
\usepackage{orcidlink}
\usepackage[T1]{fontenc}
\usepackage[mathlines]{lineno}

\usepackage{acronym}

\usepackage{etoolbox} 
\newcommand{\AckText}{}
\newcommand{\AddAck}[1]{\appto{\AckText}{#1}}

\renewcommand{\eqref}[1]{Eq.~(\ref{#1})}

\shorttitle{Sub-threshold BNS Search in LVK O4a}
\shortauthors{Niu et al.}

\begin{document}

\title{GW231109\_235456: A Sub-threshold Binary Neutron Star Merger in the LIGO–Virgo–KAGRA O4a Observing Run?}

\AddAck{This research has made use of data, software and/or web tools obtained from the
Gravitational Wave Open Science Center (\texttt{https://www.gw-openscience.org/} ), a
service of \ac{LIGO} Laboratory, the \ac{LSC} and the Virgo
Collaboration.
We especially made heavy use of the \ac{LVK} Algorithm
Library~\cite{LAL, lalsuite}.
\ac{LIGO} was constructed by the California Institute of Technology and the
Massachusetts Institute of Technology with funding from the United States
National Science Foundation (NSF) and operates under cooperative agreements
PHYS-$0757058$ and PHY-$0823459$.
In addition, the Science and Technology Facilities Council (STFC) of the United
Kingdom, the Max-Planck-Society (MPS), and the State of Niedersachsen/Germany
supported the construction of \ac{aLIGO} and construction and operation of the
GEO600 detector.
Additional support for \ac{aLIGO} was provided by the Australian Research Council.
Virgo is funded, through the European Gravitational Observatory (EGO), by the
French Centre National de Recherche Scientifique (CNRS), the Italian Istituto
Nazionale di Fisica Nucleare (INFN) and the Dutch Nikhef, with contributions by
institutions from Belgium, Germany, Greece, Hungary, Ireland, Japan, Monaco,
Poland, Portugal, Spain.
The authors are grateful for computational resources provided by
the \ac{LIGO} Lab culster at the \ac{LIGO} Laboratory and supported by
PHY-$0757058$ and PHY$-0823459$}

\AddAck{, the Pennsylvania State University's Institute
for Computational and Data Sciences (RRID:SCR_025154) and 
Pennsylvania State University RISE Core Facility (RRID:SCR_026426),
and supported by
OAC-$2103662$, PHY-$2308881$, PHY-$2011865$, OAC-$2201445$, OAC-$2018299$,
and PHY-$2207728$}

\AddAck{, and the University of Wisconsin Milwaukee Nemo and supported by PHY-$1626190$ and PHY-$2110594$}

\AddAck{.}

\author{Wanting Niu \orcidlink{0000-0003-1470-532X}}
\email{wanting.niu@ligo.org}
\affiliation{Department of Physics, The Pennsylvania State University, University Park, PA 16802, USA}
\affiliation{Institute for Gravitation and the Cosmos, The Pennsylvania State University, University Park, PA 16802, USA}

\author{Carl-Johan Haster \orcidlink{0000-0001-8040-9807}}
\affiliation{Department of Physics and Astronomy, University of Nevada, Las Vegas, 4505 South Maryland Parkway, Las Vegas, NV 89154, USA}
\affiliation{Nevada Center for Astrophysics, University of Nevada, Las Vegas, NV 89154, USA}

\author{Jolien D. E. Creighton \orcidlink{0000-0003-3600-2406}}
\affiliation{Leonard E.\ Parker Center for Gravitation, Cosmology, and Astrophysics, University of Wisconsin-Milwaukee, Milwaukee, WI 53201, USA}

\author{Shomik Adhicary \orcidlink{0009-0004-2101-5428}}
\affiliation{Department of Physics, The Pennsylvania State University, University Park, PA 16802, USA}
\affiliation{Institute for Gravitation and the Cosmos, The Pennsylvania State University, University Park, PA 16802, USA}

\author{Pratyusava Baral \orcidlink{0000-0001-6308-211X}}
\affiliation{Leonard E.\ Parker Center for Gravitation, Cosmology, and Astrophysics, University of Wisconsin-Milwaukee, Milwaukee, WI 53201, USA}

\author{Amanda Baylor \orcidlink{0000-0003-0918-0864}}
\affiliation{Leonard E.\ Parker Center for Gravitation, Cosmology, and Astrophysics, University of Wisconsin-Milwaukee, Milwaukee, WI 53201, USA}



\author{Bryce Cousins \orcidlink{0000-0002-7026-1340}}
\affiliation{Department of Physics, University of Illinois, Urbana, IL 61801 USA}
\affiliation{Department of Physics, The Pennsylvania State University, University Park, PA 16802, USA}
\affiliation{Institute for Gravitation and the Cosmos, The Pennsylvania State University, University Park, PA 16802, USA}
\AddAck{B.C. acknowledges support from the NSF Graduate Research Fellowship Program under Grant No. DGE 21-46756.}


\author{Heather Fong}
\affiliation{Department of Physics and Astronomy, University of British Columbia, Vancouver, BC, V6T 1Z4, Canada}
\affiliation{RESCEU, The University of Tokyo, Tokyo, 113-0033, Japan}
\affiliation{Graduate School of Science, The University of Tokyo, Tokyo 113-0033, Japan}

\author{Richard N. George \orcidlink{0000-0002-7797-7683}}
\affiliation{Center for Gravitational Physics, University of Texas at Austin, Austin, TX 78712, USA}



\author{Yun-Jing Huang \orcidlink{0000-0002-2952-8429}}
\affiliation{Department of Physics, The Pennsylvania State University, University Park, PA 16802, USA}
\affiliation{Institute for Gravitation and the Cosmos, The Pennsylvania State University, University Park, PA 16802, USA}

\author{Rachael Huxford}
\affiliation{Minnesota Supercomputing Institute, University of Minnesota, Minneapolis, MN 55455, USA}

\author{Prathamesh Joshi \orcidlink{0000-0002-4148-4932}}
\affiliation{Department of Physics, The Pennsylvania State University, University Park, PA 16802, USA}
\affiliation{Institute for Gravitation and the Cosmos, The Pennsylvania State University, University Park, PA 16802, USA}
\affiliation{School of Physics, Georgia Institute of Technology, Atlanta, GA 30332, USA}

\author{James Kennington \orcidlink{0000-0002-6899-3833}}
\affiliation{Department of Physics, The Pennsylvania State University, University Park, PA 16802, USA}
\affiliation{Institute for Gravitation and the Cosmos, The Pennsylvania State University, University Park, PA 16802, USA}


\author{Alvin K. Y. Li \orcidlink{0000-0001-6728-6523}}
\affiliation{RESCEU, The University of Tokyo, Tokyo, 113-0033, Japan}
\affiliation{Graduate School of Science, The University of Tokyo, Tokyo 113-0033, Japan}

\author{Ryan Magee \orcidlink{0000-0001-9769-531X}}
\affiliation{LIGO Laboratory, California Institute of Technology, Pasadena, CA 91125, USA}

\author{Duncan Meacher \orcidlink{0000-0001-5882-0368}}
\affiliation{Leonard E.\ Parker Center for Gravitation, Cosmology, and Astrophysics, University of Wisconsin-Milwaukee, Milwaukee, WI 53201, USA}

\author{Cody Messick \orcidlink{0000-0002-8230-3309}}
\affiliation{Leonard E.\ Parker Center for Gravitation, Cosmology, and Astrophysics, University of Wisconsin-Milwaukee, Milwaukee, WI 53201, USA}

\author{Soichiro Morisaki \orcidlink{0000-0002-8445-6747}}
\affiliation{Institute for Cosmic Ray Research, The University of Tokyo, 5-1-5 Kashiwanoha, Kashiwa, Chiba 277-8582, Japan}



\author{Cort Posnansky \orcidlink{0009-0009-7137-9795}}
\affiliation{Department of Physics, The Pennsylvania State University, University Park, PA 16802, USA}
\affiliation{Institute for Gravitation and the Cosmos, The Pennsylvania State University, University Park, PA 16802, USA}


\author{Surabhi Sachdev \orcidlink{0000-0002-0525-2317}}
\affiliation{School of Physics, Georgia Institute of Technology, Atlanta, GA 30332, USA}

\author{Shio Sakon \orcidlink{0000-0002-5861-3024}}
\affiliation{Department of Physics, The Pennsylvania State University, University Park, PA 16802, USA}
\affiliation{Institute for Gravitation and the Cosmos, The Pennsylvania State University, University Park, PA 16802, USA}


\author{Urja Shah \orcidlink{0000-0001-8249-7425}}
\affiliation{School of Physics, Georgia Institute of Technology, Atlanta, GA 30332, USA}

\author{Divya Singh \orcidlink{0000-0001-9675-4584}}
\affiliation{Department of Physics, University of California, Berkeley, CA 94720, USA}
\AddAck{DS acknowledges support from NSF Grant PHY-2020275(Network for Neutrinos, Nuclear Astrophysics, and Symmetries (N3AS)).}

\author{Ron Tapia}
\affiliation{Department of Physics, The Pennsylvania State University, University Park, PA 16802, USA}
\affiliation{Institute for Computational and Data Sciences, The Pennsylvania State University, University Park, PA 16802, USA}

\author{Leo Tsukada  \orcidlink{0000-0003-0596-5648}}
\affiliation{Department of Physics and Astronomy, University of Nevada, Las Vegas, 4505 South Maryland Parkway, Las Vegas, NV 89154, USA}
\affiliation{Nevada Center for Astrophysics, University of Nevada, Las Vegas, NV 89154, USA}
\AddAck{
LT acknowledges NASA 80NSSC23M0104 and the Nevada Center for Astrophysics for support.}


\author{Aaron Viets \orcidlink{0000-0002-4241-1428}}
\affiliation{Concordia University Wisconsin, Mequon, WI 53097, USA}



\author{Zach Yarbrough \orcidlink{0000-0002-9825-1136}}
\affiliation{Department of Physics and Astronomy, Louisiana State University, Baton Rouge, LA 70803, USA}

\author{Noah Zhang \orcidlink{0009-0003-3361-5538}}
\affiliation{School of Physics, Georgia Institute of Technology, Atlanta, GA 30332, USA}

\author{Chad Hanna}
\affiliation{Department of Physics, The Pennsylvania State University, University Park, PA 16802, USA}
\affiliation{Institute for Gravitation and the Cosmos, The Pennsylvania State University, University Park, PA 16802, USA}
\affiliation{Department of Astronomy and Astrophysics, The Pennsylvania State University, University Park, PA 16802, USA}
\affiliation{Institute for Computational and Data Sciences, The Pennsylvania State University, University Park, PA 16802, USA}
\AddAck{
CH Acknowledges generous support from the Eberly College of Science, the
Department of Physics, the Institute for Gravitation and the Cosmos, the
Institute for Computational and Data Sciences, and the Freed Early Career Professorship.}

\acrodef{LSC}[LSC]{LIGO Scientific Collaboration}
\acrodef{LVC}[LVC]{LIGO Scientific and Virgo Collaboration}
\acrodef{LVK}[LVK]{LIGO Scientific, Virgo and KAGRA Collaboration}
\acrodef{aLIGO}{Advanced Laser Interferometer Gravitational-Wave Observatory}
\acrodef{aVirgo}{Advanced Virgo}
\acrodef{LIGO}[LIGO]{Laser Interferometer Gravitational-Wave Observatory}
\acrodef{IFO}[IFO]{interferometer}
\acrodef{LHO}[LHO]{LIGO-Hanford}
\acrodef{LLO}[LLO]{LIGO-Livingston}
\acrodef{O2}[O2]{second observing run}
\acrodef{O1}[O1]{first observing run}
\acrodef{O3}[O3]{third observing run}
\acrodef{O3a}[O3a]{first part of the third observing run}
\acrodef{O3b}[O3b]{second part of the third observing run}
\acrodef{O4a}[O4a]{first part of the fourth observing run}
\acrodef{O4}[O4]{fourth observing run}
\acrodef{O5}[O5]{fifth observing run}

\acrodef{SSM}[SSM]{subsolar-mass}
\acrodef{BH}[BH]{black hole}
\acrodef{BBH}[BBH]{binary black hole}
\acrodef{BNS}[BNS]{binary neutron star}
\acrodef{IMBH}[IMBH]{intermediate-mass black hole}
\acrodef{NS}[NS]{neutron star}
\acrodef{BHNS}[BHNS]{black hole--neutron star binaries}
\acrodef{NSBH}[NSBH]{neutron star--black hole binary}
\acrodef{PBH}[PBH]{primordial black hole binaries}
\acrodef{CBC}[CBC]{compact binary coalescence}
\acrodef{GW}[GW]{gravitational-wave}
\acrodef{GWH}[GW]{gravitational-wave}
\acrodef{DBH}[DBH]{dispasstive black hole binaries}
\acrodef{DNS}[DNS]{double neutron star}
\acrodef{PK}[PK]{Post-Keplerian}

\acrodef{SNR}[SNR]{signal-to-noise ratio}
\acrodef{FAR}[FAR]{false alarm rate}
\acrodef{FAP}[FAP]{false alarm probability}
\acrodef{PSD}[PSD]{power spectral density}
\acrodef{GR}[GR]{general relativity}
\acrodef{NR}[NR]{numerical relativity}
\acrodef{PN}[PN]{post-Newtonian}
\acrodef{EOB}[EOB]{effective-one-body}
\acrodef{ROM}[ROM]{reduced-order model}
\acrodef{IMR}[IMR]{inspiral--merger--ringdown}
\acrodef{EOS}[EoS]{equation of state}
\acrodef{FF}[FF]{fitting factor}
\acrodef{FT}[FT]{Fourier Transform}
\acrodef{GRB}[GRB]{gamma-ray burst}
\acrodef{ROQ}[ROQ]{reduced-order quadrature}
\acrodef{CI}[CI]{credible interval}

\acrodef{LAL}[LAL]{LIGO Algorithm Library}
\acrodef{GWTC}[GWTC]{Gravitational Wave Transient Catalog}

\newcommand{\PN}[0]{\ac{PN}\xspace}
\newcommand{\BBH}[0]{\ac{BBH}\xspace}
\newcommand{\BNS}[0]{\ac{BNS}\xspace}
\newcommand{\BH}[0]{\ac{BH}\xspace}
\newcommand{\NR}[0]{\ac{NR}\xspace}
\newcommand{\GW}[0]{\ac{GW}\xspace}
\newcommand{\SNR}[0]{\ac{SNR}\xspace}
\newcommand{\SSM}[0]{\ac{SSM}\xspace}
\newcommand{\aLIGO}[0]{\ac{aLIGO}\xspace}
\newcommand{\PSD}[0]{\ac{PSD}\xspace}
\newcommand{\GR}[0]{\ac{GR}\xspace}
\newcommand{\EOS}[0]{\ac{EOS}\xspace}
\newcommand{\LVC}[0]{\ac{LVC}\xspace}


\newcommand{\GSTLAL}{GstLAL\xspace}
\newcommand{\IMRPHENOMD}{IMRPhenomD\xspace}
\newcommand{\MANIFOLD}{{\fontfamily{qcr}\selectfont manifold}\xspace}
\newcommand{\SBANK}{{\fontfamily{qcr}\selectfont SBank}\xspace}

\newcommand\hmm[1]{\ifnum\ifhmode\spacefactor\else2000\fi>1500 \uppercase{#1}\else#1\fi}

\newcommand{\IGWNALERT}{\texttt{igwn-alert}\xspace}

\newcommand{\MDCSTART}{5 Jan. 2020 15:59:42}
\newcommand{\MDCEND}{14 Feb. 2020 15:59:42}

\newcommand{\TOTALTEMPLATES}{\ensuremath{1.8 \times 10^6}}
\newcommand{\CHECKERBOARDTEMPLATES}{\ensuremath{9 \times 10^5}}
\newcommand{\NUMSVDBANKS}{\ensuremath{\sim 1000}}
\newcommand{\TEMPLATESPERSUBBANK}{\ensuremath{\sim 500}}
\newcommand{\NUMSUBBANKSPERSVD}{\ensuremath{2}}
\newcommand{\SVDTOLERANCE}{\ensuremath{99.999\%}}
\newcommand{\PSDFFTLENGTH}{\ensuremath{4~\mathrm{seconds}}}
\newcommand{\FRAMELENGTH}{\ensuremath{1~\mathrm{second}}}
\newcommand{\BUFFERBLOCKSIZE}{\ensuremath{4096~\mathrm{bytes}}}
\newcommand{\FIRSTRIDE}{\ensuremath{0.25~\mathrm{seconds}}}
\newcommand{\TRIGGERSNRTHRESHOLD}{\ensuremath{4.0}}
\newcommand{\COINCTHRESHOLD}{\ensuremath{0.005~\mathrm{seconds}}}
\newcommand{\HTGATETHRESHOLDMIN}{\ensuremath{15.0}}
\newcommand{\HTGATETHRESHOLDMAX}{\ensuremath{100.0}}
\newcommand{\HTGATEMCHIRPMIN}{\ensuremath{0.8}}
\newcommand{\HTGATEMCHIRPMAX}{\ensuremath{45.0}}
\newcommand{\HTGATEMIN}{\ensuremath{\sim 15}}
\newcommand{\HTGATEMAX}{\ensuremath{\sim 325}}
\newcommand{\LRSNAPSHOT}{\ensuremath{4}}
\newcommand{\LRCOMPRESSION}{\ensuremath{0.003}}
\newcommand{\FARTRIALSFACTOR}{\ensuremath{2}}
\newcommand{\UPLOADCADENCE}{\ensuremath{4}}
\newcommand{\UPLOADCADENCEMDCTWELVE}{\ensuremath{2}}
\newcommand{\UPLOADDT}{\ensuremath{0.2}}
\newcommand{\SINGLESPENALTYMDCELEVEN}{\ensuremath{12}}
\newcommand{\SINGLESPENALTYOFOUR}{\ensuremath{13}}
\newcommand{\XISQMISMATCHRANGE}{\ensuremath{0.1-10\%}}

\newcommand{\TESTSUITECOINCWINDOW}{\ensuremath{\pm 1}}
\newcommand{\INJSNRFLOW}{\ensuremath{10.0}}
\newcommand{\INJSNRFHI}{\ensuremath{1600.0}}

\newcommand{\VT}{\ensuremath{\langle VT \rangle}}
\newcommand{\SPINZ}{\ensuremath{s_{i,z}}}
\newcommand{\CHIP}{\ensuremath{\chi_p}}
\newcommand{\MCHIRP}{\ensuremath{\mathcal{M}_c}\xspace}
\newcommand{\CHIEFF}{\ensuremath{\chi_{\mathrm{eff}}}}
\newcommand{\PASTRO}{\ensuremath{p(\mathrm{astro})}}
\newcommand{\MSUN}{\ensuremath{M_{\odot}}}
\newcommand{\TEND}{\ensuremath{t_{\mathrm{end}}}}

\newcommand{\TOTALINJECTIONS}{$5\times10^4$}
\newcommand{\BNSMAXZ}{\ensuremath{0.15}}
\newcommand{\NSBHMAXZ}{\ensuremath{0.25}}
\newcommand{\BBHMAXZ}{\ensuremath{1.9}}
\newcommand{\MDCDURATION}{\ensuremath{3.456\times10^6}}
\newcommand{\INJECTIONSPACING}{\ensuremath{\sim40}}

\newcommand{\PASTROTHRESHOLD}{\ensuremath{0.50}}

\newcommand{\DECISIVESNRTHRESH}{\ensuremath{8.0}}
\newcommand{\NETWORKSNRTHRESH}{\ensuremath{10.0}}

\newcommand{\ALLABOVEDECSNRTHRESH}{\ensuremath{1522}}
\newcommand{\ALLINBANKABOVEDECSNRTHRESH}{\ensuremath{1457}}
\newcommand{\BBHINBANKABOVEDECSNRTHRESH}{\ensuremath{597}}
\newcommand{\BNSINBANKABOVEDECSNRTHRESH}{\ensuremath{482}}
\newcommand{\NSBHINBANKABOVEDECSNRTHRESH}{\ensuremath{378}}

\newcommand{\HIGHFARTHRESH}{$1$ per hour}
\newcommand{\LOWFARTHRESH}{$2$ per day}
\newcommand{\ONEPERHOUR}{\ensuremath{2.78\times10^{-4}~\mathrm{Hz}}}
\newcommand{\TWOPERDAY}{\ensuremath{2.31\times10^{-5}~\mathrm{Hz}}}
\newcommand{\ONEPERMONTH}{\ensuremath{3.85\times10^{-7}~\mathrm{Hz}}}
\newcommand{\TWOPERYEAR}{\ensuremath{3.16\times10^{-8}~\mathrm{Hz}}}

\newcommand{\ALLINBANKEFFICIENCY}[1]{%
	\IfEqCase{#1}{%
		{ONEPERHOUR}{\ensuremath{0.87}}%
		{TWOPERDAY}{\ensuremath{0.84}}%
		{ONEPERMONTH}{\ensuremath{0.78}}%
		{TWOPERYEAR}{\ensuremath{0.74}}%
	}[\PackageError{ALLINBANKEFFICIENCY}{Undefined option: #1}{}]
}%

\newcommand{\BBHINBANKEFFICIENCY}[1]{%
	\IfEqCase{#1}{%
		{ONEPERHOUR}{\ensuremath{0.87}}%
		{TWOPERDAY}{\ensuremath{0.84}}%
		{ONEPERMONTH}{\ensuremath{0.77}}%
		{TWOPERYEAR}{\ensuremath{0.71}}%
	}[\PackageError{BBHINBANKEFFICIENCY}{Undefined option: #1}{}]
}%

\newcommand{\BNSINBANKEFFICIENCY}[1]{%
	\IfEqCase{#1}{%
		{ONEPERHOUR}{\ensuremath{0.95}}%
		{TWOPERDAY}{\ensuremath{0.95}}%
		{ONEPERMONTH}{\ensuremath{0.89}}%
		{TWOPERYEAR}{\ensuremath{0.86}}%
	}[\PackageError{BNSINBANKEFFICIENCY}{Undefined option: #1}{}]
}%

\newcommand{\NSBHINBANKEFFICIENCY}[1]{%
	\IfEqCase{#1}{%
		{ONEPERHOUR}{\ensuremath{0.77}}%
		{TWOPERDAY}{\ensuremath{0.71}}%
		{ONEPERMONTH}{\ensuremath{0.65}}%
		{TWOPERYEAR}{\ensuremath{0.62}}%
	}[\PackageError{NSBHINBANKEFFICIENCY}{Undefined option: #1}{}]
}%

\newcommand{\ALLABOVEDECSNRTHRESHMDCTWELVE}{\ensuremath{653}}
\newcommand{\ALLINBANKABOVEDECSNRTHRESHMDCTWELVE}{\ensuremath{621}}
\newcommand{\BBHINBANKABOVEDECSNRTHRESHMDCTWELVE}{\ensuremath{243}}
\newcommand{\BNSINBANKABOVEDECSNRTHRESHMDCTWELVE}{\ensuremath{209}}
\newcommand{\NSBHINBANKABOVEDECSNRTHRESHMDCTWELVE}{\ensuremath{169}}

\newcommand{\ALLINBANKEFFICIENCYMDCTWELVE}[1]{%
	\IfEqCase{#1}{%
		{ONEPERHOUR}{\ensuremath{0.88}}%
		{TWOPERDAY}{\ensuremath{0.86}}%
		{ONEPERMONTH}{\ensuremath{0.83}}%
		{TWOPERYEAR}{\ensuremath{0.81}}%
	}[\PackageError{ALLINBANKEFFICIENCYMDCTWELVE}{Undefined option: #1}{}]
}%

\newcommand{\BBHINBANKEFFICIENCYMDCTWELVE}[1]{%
	\IfEqCase{#1}{%
		{ONEPERHOUR}{\ensuremath{0.92}}%
		{TWOPERDAY}{\ensuremath{0.90}}%
		{ONEPERMONTH}{\ensuremath{0.87}}%
		{TWOPERYEAR}{\ensuremath{0.86}}%
	}[\PackageError{BBHINBANKEFFICIENCYMDCTWELVE}{Undefined option: #1}{}]
}%

\newcommand{\BNSINBANKEFFICIENCYMDCTWELVE}[1]{%
	\IfEqCase{#1}{%
		{ONEPERHOUR}{\ensuremath{0.95}}%
		{TWOPERDAY}{\ensuremath{0.93}}%
		{ONEPERMONTH}{\ensuremath{0.91}}%
		{TWOPERYEAR}{\ensuremath{0.88}}%
	}[\PackageError{BNSINBANKEFFICIENCYMDCTWELVE}{Undefined option: #1}{}]
}%

\newcommand{\NSBHINBANKEFFICIENCYMDCTWELVE}[1]{%
	\IfEqCase{#1}{%
		{ONEPERHOUR}{\ensuremath{0.74}}%
		{TWOPERDAY}{\ensuremath{0.72}}%
		{ONEPERMONTH}{\ensuremath{0.68}}%
		{TWOPERYEAR}{\ensuremath{0.66}}%
	}[\PackageError{NSBHINBANKEFFICIENCYMDCTWELVE}{Undefined option: #1}{}]
}%

\newcommand{\MEAN}[1]{%
	\IfEqCase{#1}{%
		{MASSRATIO}{\ensuremath{1.39}}%
		{MCHIRP}{\ensuremath{0.15}}%
		{SPIN1Z}{\ensuremath{7.27}}%
		{SPIN2Z}{\ensuremath{2.81}}%
		{CHIEFF}{\ensuremath{5.77}}%
		{ENDTIME}{\ensuremath{6.23}}%
	}[\PackageError{MEAN}{Undefined option: #1}{}]
}%

\newcommand{\STDEV}[1]{%
	\IfEqCase{#1}{%
		{MASSRATIO}{\ensuremath{2.86}}%
		{MCHIRP}{\ensuremath{0.45}}%
		{SPIN1Z}{\ensuremath{285}}%
		{SPIN2Z}{\ensuremath{183}}%
		{CHIEFF}{\ensuremath{252}}%
		{ENDTIME}{\ensuremath{30.22}}%
	}[\PackageError{STDEV}{Undefined option: #1}{}]
}%

\newcommand{\BNSMCHIRPMEAN}{\ensuremath{2.06\times10^{-4}}}
\newcommand{\BNSMCHIRPSTDEV}{\ensuremath{8.33\times10^{-4}}}

\newcommand{\NSBHMCHIRPMEAN}{\ensuremath{-2.14\times10^{-4}}}
\newcommand{\NSBHMCHIRPSTDEV}{\ensuremath{6.26\times10^{-3}}}

\newcommand{\BBHMCHIRPMEAN}{\ensuremath{1.54\times10^{-1}}}
\newcommand{\BBHMCHIRPSTDEV}{\ensuremath{4.53\times10^{-1}}}

\newcommand{\BNSENDTIMEMEAN}{\ensuremath{-0.90}}
\newcommand{\BNSENDTIMESTDEV}{\ensuremath{18.0}}

\newcommand{\NSBHENDTIMEMEAN}{\ensuremath{18.7}}
\newcommand{\NSBHENDTIMESTDEV}{\ensuremath{59.3}}

\newcommand{\BBHENDTIMEMEAN}{\ensuremath{6.03}}
\newcommand{\BBHENDTIMESTDEV}{\ensuremath{11.3}}

\newcommand{\QFIFTY}[1]{%
	\IfEqCase{#1}{%
		{MASSRATIO}{\ensuremath{0.45}}%
		{MCHIRP}{\ensuremath{0.007}}%
		{SPIN1Z}{\ensuremath{1.24}}%
		{SPIN2Z}{\ensuremath{1.72}}%
		{CHIEFF}{\ensuremath{1.34}}%
		{ENDTIME}{\ensuremath{3.8}}%
	}[\PackageError{QFIFTY}{Undefined option: #1}{}]
}%

\newcommand{\QSEVENTYFIVE}[1]{%
	\IfEqCase{#1}{%
		{MASSRATIO}{\ensuremath{1.67}}%
		{MCHIRP}{\ensuremath{0.33}}%
		{SPIN1Z}{\ensuremath{3.82}}%
		{SPIN2Z}{\ensuremath{5.33}}%
		{CHIEFF}{\ensuremath{3.71}}%
		{ENDTIME}{\ensuremath{9.78}}%
	}[\PackageError{QSEVENTYFIVE}{Undefined option: #1}{}]
}%

\newcommand{\QNINETY}[1]{%
	\IfEqCase{#1}{%
		{MASSRATIO}{\ensuremath{4.97}}%
		{MCHIRP}{\ensuremath{0.73}}%
		{SPIN1Z}{\ensuremath{13.8}}%
		{SPIN2Z}{\ensuremath{17.5}}%
		{CHIEFF}{\ensuremath{10.8}}%
		{ENDTIME}{\ensuremath{25.6}}%
	}[\PackageError{QNINETY}{Undefined option: #1}{}]
}%

\newcommand{\GPCYRS}{\ensuremath{\mathrm{Gpc}^3\mathrm{yrs}}}
\newcommand{\INJECTEDVT}[1]{%
	\IfEqCase{#1}{%
		{BNS}{\ensuremath{1.08\times10^{-1}}}%
		{NSBH}{\ensuremath{4.34\times10^{-1}}}%
		{BBH}{\ensuremath{29.1}}%
	}[\PackageError{INJECTEDVT}{Undefined option: #1}{}]
}%

\newcommand{\VTTWOPERDAY}[1]{%
	\IfEqCase{#1}{%
		{BNS}{\ensuremath{3.49\times10^{-4}}}%
		{NSBH}{\ensuremath{8.08\times10^{-4}}}%
		{BBH}{\ensuremath{1.23\times10^{-1}}}%
	}[\PackageError{VTTWOPERDAY}{Undefined option: #1}{}]
}%

\newcommand{\VTNETSNR}[1]{%
	\IfEqCase{#1}{%
		{BNS}{\ensuremath{4.41\times10^{-4}}}%
		{NSBH}{\ensuremath{1.59\times10^{-3}}}%
		{BBH}{\ensuremath{1.52\times10^{-1}}}%
	}[\PackageError{VTNETSNR}{Undefined option: #1}{}]
}%

\newcommand{\VTDECSNR}[1]{%
	\IfEqCase{#1}{%
		{BNS}{\ensuremath{1.47\times10^{-4}}}%
		{NSBH}{\ensuremath{4.98\times10^{-3}}}%
		{BBH}{\ensuremath{5.46\times10^{-2}}}%
	}[\PackageError{VTDECSNR}{Undefined option: #1}{}]
}%


\newcommand{\SEARCHEDAREAQFIFTY}[1]{%
	\IfEqCase{#1}{%
		{ALL}{\ensuremath{271}}%
		{TRIPLE}{\ensuremath{31.9}}%
		{DOUBLE}{\ensuremath{301}}%
		{SINGLE}{\ensuremath{3150}}%
	}[\PackageError{SEARCHEDAREAQFIFTY}{Undefined option: #1}{}]
}%

\newcommand{\SEARCHEDAREAQSEVENTYFIVE}[1]{%
	\IfEqCase{#1}{%
		{ALL}{\ensuremath{1080}}%
		{TRIPLE}{\ensuremath{140}}%
		{DOUBLE}{\ensuremath{893}}%
		{SINGLE}{\ensuremath{10,400}}%
	}[\PackageError{SEARCHEDAREAQSEVENTYFIVE}{Undefined option: #1}{}]
}%

\newcommand{\SEARCHEDAREAQNINETY}[1]{%
	\IfEqCase{#1}{%
		{ALL}{\ensuremath{3910}}%
		{TRIPLE}{\ensuremath{357}}%
		{DOUBLE}{\ensuremath{2470}}%
		{SINGLE}{\ensuremath{18,400}}%
	}[\PackageError{SEARCHEDAREAQNINETY}{Undefined option: #1}{}]
}%

\newcommand{\SEARCHEDPROBQFIFTY}[1]{%
	\IfEqCase{#1}{%
		{ALL}{\ensuremath{0.53}}%
		{TRIPLE}{\ensuremath{0.58}}%
		{DOUBLE}{\ensuremath{0.52}}%
		{SINGLE}{\ensuremath{0.59}}%
	}[\PackageError{SEARCHEDPROBQFIFTY}{Undefined option: #1}{}]
}%

\newcommand{\SEARCHEDPROBQSEVENTYFIVE}[1]{%
	\IfEqCase{#1}{%
		{ALL}{\ensuremath{0.79}}%
		{TRIPLE}{\ensuremath{0.84}}%
		{DOUBLE}{\ensuremath{0.77}}%
		{SINGLE}{\ensuremath{0.78}}%
	}[\PackageError{SEARCHEDPROBQSEVENTYFIVE}{Undefined option: #1}{}]
}%

\newcommand{\SEARCHEDPROBQNINETY}[1]{%
	\IfEqCase{#1}{%
		{ALL}{\ensuremath{0.93}}%
		{TRIPLE}{\ensuremath{0.96}}%
		{DOUBLE}{\ensuremath{0.92}}%
		{SINGLE}{\ensuremath{0.92}}%
	}[\PackageError{SEARCHEDPROBQNINETY}{Undefined option: #1}{}]
}%

\newcommand{\BNSTOBNS}{\ensuremath{90.3\%}}
\newcommand{\BNSTONSBH}{\ensuremath{9.7\%}}

\newcommand{\NSBHTONSBH}{\ensuremath{64.1\%}}
\newcommand{\NSBHTOBBH}{\ensuremath{33.8\%}}
\newcommand{\NSBHTOBNS}{\ensuremath{2.10\%}}

\newcommand{\BBHTOBBH}{\ensuremath{100\%}}

\newcommand{\TERRTOTERR}{\ensuremath{2.60\%}}
\newcommand{\TERRTOBBH}{\ensuremath{68.8\%}}
\newcommand{\TERRTONSBH}{\ensuremath{19.5\%}}
\newcommand{\TERRTOBNS}{\ensuremath{9.10\%}}

\newcommand{\BNSTOBNSMDCTWELVE}{\ensuremath{79.8\%}}
\newcommand{\BNSTONSBHMDCTWELVE}{\ensuremath{20.2\%}}

\newcommand{\NSBHTONSBHMDCTWELVE}{\ensuremath{92.1\%}}
\newcommand{\NSBHTOBBHMDCTWELVE}{\ensuremath{6.83\%}}
\newcommand{\NSBHTOBNSMDCTWELVE}{\ensuremath{1.02\%}}

\newcommand{\BBHTOBBHMDCTWELVE}{\ensuremath{99.5\%}}
\newcommand{\BBHTONSBHMDCTWELVE}{\ensuremath{0.05\%}}

\newcommand{\TERRTOBBHMDCTWELVE}{\ensuremath{76.2\%}}
\newcommand{\TERRTONSBHMDCTWELVE}{\ensuremath{23.8\%}}

\newcommand{\OTHREEOPA}{\ensuremath{1.2}} 

\newcommand{\MDCGWIFOS}[1]{%
	\IfEqCase{#1}{%
		{GW200112}{L1}%
		{GW200115}{H1L1}%
		{GW200128}{H1L1}%
		{GW200129}{H1L1V1}%
		{GW200202}{H1L1}%
		{GW200208q}{H1L1}%
		{GW200208am}{H1L1}%
		{GW200209}{H1L1}%
		{GW200210}{H1L1}%
	}[\PackageError{MDCGWIFOS}{Undefined option: #1}{}]
}%

\newcommand{\MDCGWSNR}[1]{%
	\IfEqCase{#1}{%
		{GW200112}{\ensuremath{18.46}}%
		{GW200115}{\ensuremath{11.48}}%
		{GW200128}{\ensuremath{9.98}}%
		{GW200129}{\ensuremath{26.30}}%
		{GW200202}{\ensuremath{11.09}}%
		{GW200208q}{\ensuremath{10.56}}%
		{GW200208am}{\ensuremath{8.00}}%
		{GW200209}{\ensuremath{9.96}}%
		{GW200210}{\ensuremath{9.28}}%
	}[\PackageError{MDCGWSNR}{Undefined option: #1}{}]
}%

\newcommand{\MDCGWFAR}[1]{%
	\IfEqCase{#1}{%
		{GW200112}{\ensuremath{1.01\times10^{-7}}}%
		{GW200115}{\ensuremath{2.55\times10^{-4}}}%
		{GW200128}{\ensuremath{1.44\times10^{-4}}}%
		{GW200129}{\ensuremath{1.78\times10^{-17}}}%
		{GW200202}{\ensuremath{1.69\times10^{-2}}}%
		{GW200208q}{\ensuremath{4.92\times10^{-5}}}%
		{GW200208am}{\ensuremath{2.02\times10^{3}}}%
		{GW200209}{\ensuremath{1.20}}%
		{GW200210}{\ensuremath{3.64\times10^{3}}}%
	}[\PackageError{MDCGWFAR}{Undefined option: #1}{}]
}%

\newcommand{\MDCGWPASTRO}[1]{%
	\IfEqCase{#1}{%
		{GW200112}{\ensuremath{>0.99}}%
		{GW200115}{\ensuremath{>0.99}}%
		{GW200128}{\ensuremath{>0.99}}%
		{GW200129}{\ensuremath{>0.99}}%
		{GW200202}{\ensuremath{>0.99}}%
		{GW200208q}{\ensuremath{>0.99}}%
		{GW200208am}{\ensuremath{0.48}}%
		{GW200209}{\ensuremath{>0.99}}%
		{GW200210}{0.27}%
	}[\PackageError{MDCGWPASTRO}{Undefined option: #1}{}]
}%

\newcommand{\MDCGWMCHIRP}[1]{%
	\IfEqCase{#1}{%
		{GW200112}{\ensuremath{33.37~M_{\odot}}}%
		{GW200115}{\ensuremath{2.58~M_{\odot}}}%
		{GW200128}{\ensuremath{50.74~M_{\odot}}}%
		{GW200129}{\ensuremath{30.66~M_{\odot}}}%
		{GW200202}{\ensuremath{8.15~M_{\odot}}}%
		{GW200208q}{\ensuremath{34.50~M_{\odot}}}%
		{GW200208am}{\ensuremath{66.59~M_{\odot}}}%
		{GW200209}{\ensuremath{39.45~M_{\odot}}}%
		{GW200210}{\ensuremath{7.89~M_{\odot}}}%
	}[\PackageError{MDCGWMCHIRP}{Undefined option: #1}{}]
}%

\newcommand{\OTHREEGWIFOS}[1]{%
	\IfEqCase{#1}{%
		{GW200112}{L1}%
		{GW200115}{H1L1}%
		{GW200128}{--}%
		{GW200129}{H1L1V1}%
		{GW200202}{--}%
		{GW200208q}{--}%
		{GW200208am}{--}%
		{GW200209}{--}%
		{GW200210}{--}%
	}[\PackageError{OTHREEGWIFOS}{Undefined option: #1}{}]
}%

\newcommand{\OTHREEGWSNR}[1]{%
	\IfEqCase{#1}{%
		{GW200112}{\ensuremath{18.79}}%
		{GW200115}{\ensuremath{11.42}}%
		{GW200128}{--}%
		{GW200129}{\ensuremath{26.61}}%
		{GW200202}{--}%
		{GW200208q}{--}%
		{GW200208am}{--}%
		{GW200209}{--}%
		{GW200210}{--}%
	}[\PackageError{OTHREEGWSNR}{Undefined option: #1}{}]
}%

\newcommand{\OTHREEGWFAR}[1]{%
	\IfEqCase{#1}{%
		{GW200112}{\ensuremath{4.05\times10^{-4}}}%
		{GW200115}{\ensuremath{6.61\times10^{-4}}}%
		{GW200128}{\ensuremath{> \OTHREEOPA{}}}%
		{GW200129}{\ensuremath{2.11\times10^{-24}}}%
		{GW200202}{\ensuremath{> \OTHREEOPA{}}}%
		{GW200208q}{--}%
		{GW200208am}{--}%
		{GW200209}{--}%
		{GW200210}{--}%
	}[\PackageError{OTHREEGWFAR}{Undefined option: #1}{}]
}%

\newcommand{\OTHREEGWPASTRO}[1]{%
	\IfEqCase{#1}{%
		{GW200112}{\ensuremath{>0.99}}%
		{GW200115}{\ensuremath{>0.99}}%
		{GW200128}{--}%
		{GW200129}{\ensuremath{>0.99}}%
		{GW200202}{--}%
		{GW200208q}{--}%
		{GW200208am}{--}%
		{GW200209}{--}%
		{GW200210}{--}%
	}[\PackageError{OTHREEGWPASTRO}{Undefined option: #1}{}]
}%

\newcommand{\OTHREEGWMCHIRP}[1]{%
	\IfEqCase{#1}{%
		{GW200112}{\ensuremath{35.37~M_{\odot}}}%
		{GW200115}{\ensuremath{2.57~M_{\odot}}}%
		{GW200128}{--}%
		{GW200129}{\ensuremath{32.74~M_{\odot}}}%
		{GW200202}{--}%
		{GW200208q}{--}%
		{GW200208am}{--}%
		{GW200209}{--}%
		{GW200210}{--}%
	}[\PackageError{OTHREEGWMCHIRP}{Undefined option: #1}{}]
}

\newcommand{\OTHREERETRACTIONS}{23}
\newcommand{\OTHREEGSTLALRETRACTIONS}{15}

\newcommand{\RETRACTIONFAR}{\ensuremath{1.67~\mathrm{per}~\mathrm{year}}}
\newcommand{\RETRACTIONSNR}{\ensuremath{14.5}}
\newcommand{\MDCRETRACTIONFARTHRESH}{one per year}

\newcommand{\BANKMASSLOW}{\ensuremath{1.0~M_{\odot}}}
\newcommand{\BANKMASSHIGH}{\ensuremath{200~M_{\odot}}}

\newcommand{\BHMASSLOW}{\ensuremath{3.0~M_{\odot}}}
\newcommand{\NSMASSLOW}{\ensuremath{1.0~M_{\odot}}}
\newcommand{\NSMASSHIGH}{\ensuremath{3.0 M_{\odot}}}
\newcommand{\TOTALMASSHIGH}{\ensuremath{400.0~M_{\odot}}}
\newcommand{\MASSRATIOHIGH}{\ensuremath{20}}

\newcommand{\NSSPIN}{\ensuremath{0.05}}
\newcommand{\BHSPIN}{\ensuremath{0.99}}
\newcommand{\CHIPBOUND}{\ensuremath{1\times10^{-3}}}

\newcommand{\MCHIRPBOUNDARY}{\ensuremath{1.73~M_{\odot}}}
\newcommand{\LOWMCHIRPWAVEFORM}{\texttt{TaylorF2}}
\newcommand{\HIGHMCHIRPWAVEFORM}{\texttt{SEOBNRv4}}

\newcommand{\MDCELEVENLOFARLATENCY}{\ensuremath{14.58}}
\newcommand{\MDCELEVENHIFARLATENCY}{\ensuremath{10.30}}

\newcommand{\MDCTWELVELOFARLATENCY}{\ensuremath{12.04}}
\newcommand{\MDCTWELVEHIFARLATENCY}{\ensuremath{9.86}}

\newcommand{\logTenBFHSoverLS}{\ensuremath 0.33}

\newcommand{\LowSpinPriorUpperBound}{\ensuremath 0.05}
\newcommand{\HighSpinPriorUpperBound}{\ensuremath 0.40}

\newcommand{\LSMassoneSourceZeroPercentile}{\ensuremath 1.40}
\newcommand{\LSMassoneSourceNinetyPercentile}{\ensuremath 1.65}

\newcommand{\LSMasstwoSourceTenPercentile}{\ensuremath 1.27}
\newcommand{\LSMasstwoSourceHundredPercentile}{\ensuremath 1.49}

\newcommand{\LSChirpMassSourceFivePercentile}{\ensuremath 0.018}
\newcommand{\LSChirpMassSourceFiftyPercentile}{\ensuremath 1.260}
\newcommand{\LSChirpMassSourceNinetyFivePercentile}{\ensuremath 0.019}

\newcommand{\LSChirpMassFivePercentile}{\ensuremath 0.0003}
\newcommand{\LSChirpMassFiftyPercentile}{\ensuremath 1.3063}
\newcommand{\LSChirpMassNinetyFivePercentile}{\ensuremath 0.0003}

\newcommand{\LSMassRatioTenPercentile}{\ensuremath 0.77}
\newcommand{\LSMassRatioHundredPercentile}{\ensuremath 1.00}

\newcommand{\LSTotalMassSourceFivePercentile}{\ensuremath 0.04}
\newcommand{\LSTotalMassSourceFiftyPercentile}{\ensuremath 2.90}
\newcommand{\LSTotalMassSourceNinetyFivePercentile}{\ensuremath 0.05}

\newcommand{\LSChiEffFivePercentile}{\ensuremath 0.015}
\newcommand{\LSChiEffFiftyPercentile}{\ensuremath 0.014}
\newcommand{\LSChiEffNinetyFivePercentile}{\ensuremath 0.016}

\newcommand{\LSLuminosityDistanceFivePercentile}{\ensuremath 70}
\newcommand{\LSLuminosityDistanceFiftyPercentile}{\ensuremath 169}
\newcommand{\LSLuminosityDistanceNinetyFivePercentile}{\ensuremath 72}

\newcommand{\LSViewingAngleFivePercentile}{\ensuremath 10}
\newcommand{\LSViewingAngleNinetyFivePercentile}{\ensuremath 65}
\newcommand{\LSViewingAngleZeroPercentile}{\ensuremath 0}
\newcommand{\LSViewingAngleNinetyPercentile}{\ensuremath 60}

\newcommand{\LSLambdaSymmetricZeroPercentile}{\ensuremath 0}
\newcommand{\LSLambdaSymmetricNinetyPercentile}{\ensuremath 2700}

\newcommand{\LSLambdaTildeZeroPercentile}{\ensuremath 0}
\newcommand{\LSLambdaTildeNinetyPercentile}{\ensuremath 2500}
\newcommand{\LSLambdaTildeNinetyPercentileReweightBinaryLove}{\ensuremath 3700}

\newcommand{\LSlogTenBFBLoverBH}{\ensuremath -0.23}
\newcommand{\LSlogTenBFBLoverfL}{\ensuremath 0.39}

\newcommand{\LSLocalizationAreaNinteyPercent}{\ensuremath 450}


\newcommand{\HSMassoneSourceZeroPercentile}{\ensuremath 1.40}
\newcommand{\HSMassoneSourceNinetyPercentile}{\ensuremath 2.24}

\newcommand{\HSMasstwoSourceTenPercentile}{\ensuremath 0.97}
\newcommand{\HSMasstwoSourceHundredPercentile}{\ensuremath 1.49}

\newcommand{\HSChirpMassSourceFivePercentile}{\ensuremath 0.018}
\newcommand{\HSChirpMassSourceFiftyPercentile}{\ensuremath 1.261}
\newcommand{\HSChirpMassSourceNinetyFivePercentile}{\ensuremath 0.018}

\newcommand{\HSChirpMassFivePercentile}{\ensuremath 0.0005}
\newcommand{\HSChirpMassFiftyPercentile}{\ensuremath 1.3067}
\newcommand{\HSChirpMassNinetyFivePercentile}{\ensuremath 0.0006}

\newcommand{\HSMassRatioTenPercentile}{\ensuremath 0.43}
\newcommand{\HSMassRatioHundredPercentile}{\ensuremath 1.00}

\newcommand{\HSTotalMassSourceFivePercentile}{\ensuremath 0.07}
\newcommand{\HSTotalMassSourceFiftyPercentile}{\ensuremath 2.95}
\newcommand{\HSTotalMassSourceNinetyFivePercentile}{\ensuremath 0.38}

\newcommand{\HSChiEffFivePercentile}{\ensuremath 0.035}
\newcommand{\HSChiEffFiftyPercentile}{\ensuremath 0.051}
\newcommand{\HSChiEffNinetyFivePercentile}{\ensuremath 0.103}

\newcommand{\HSLuminosityDistanceFivePercentile}{\ensuremath 69}
\newcommand{\HSLuminosityDistanceFiftyPercentile}{\ensuremath 165}
\newcommand{\HSLuminosityDistanceNinetyFivePercentile}{\ensuremath 70}

\newcommand{\HSViewingAngleFivePercentile}{\ensuremath 10}
\newcommand{\HSViewingAngleNinetyFivePercentile}{\ensuremath 66}
\newcommand{\HSViewingAngleZeroPercentile}{\ensuremath 0}
\newcommand{\HSViewingAngleNinetyPercentile}{\ensuremath 60}

\newcommand{\HSLambdaSymmetricZeroPercentile}{\ensuremath 0}
\newcommand{\HSLambdaSymmetricNinetyPercentile}{\ensuremath 3000}

\newcommand{\HSLambdaTildeZeroPercentile}{\ensuremath 5}
\newcommand{\HSLambdaTildeNinetyPercentile}{\ensuremath 2791}
\newcommand{\HSLambdaTildeNinetyPercentileReweightBinaryLove}{\ensuremath 4500}

\newcommand{\HSLocalizationAreaNinteyPercent}{\ensuremath 430}

\newcommand{\HSlogTenBFBLoverBH}{\ensuremath 0.09}
\newcommand{\HSlogTenBFBLoverfL}{\ensuremath 0.12}

\newcommand{\FalseAlarmRateSigTrigger}{\ensuremath 0.017}
\newcommand{\FalseAlarmRateGWTCFOURTrials}{\ensuremath 0.083}
\newcommand{\NetworkSNRSigTrigger}{\ensuremath 9.7}
\newcommand{\HanfordSNRSigTrigger}{\ensuremath 7.1}
\newcommand{\LivingstonSNRSigTrigger}{\ensuremath 6.5}
\newcommand{\InstrumentsSigTrigger}{HL}
\newcommand{\GPSTimeSigTrigger}{\ensuremath 1383609314.057}
\newcommand{\UTCTimeSigTrigger}{\ensuremath 9 Nov 2023, 23:54:56.057}
\newcommand{\ChirpMassSigTrigger}{\ensuremath 1.306}
\newcommand{\PrimaryMassSigTrigger}{\ensuremath 1.78}
\newcommand{\SecondaryMassSigTrigger}{\ensuremath 1.27}

\newcommand{\FirstHalfFourthObservationStartTimeUTC}{2023 May 24 15:00:00}
\newcommand{\FirstHalfFourthObservationEndTimeUTC}{2024 January 16 16:00:00}

\newcommand{\FirstHalfFourthObservationChunkOneStartTimeUTC}{2023 May 24 15:00:00 UTC}
\newcommand{\FirstHalfFourthObservationChunkOneEndTimeUTC}{2023 Jun 06 14:30:34 UTC}
\newcommand{\DNSVTUnit}{\,\mathrm{Gpc}^{3}\,\mathrm{yr}^{1}}
\newcommand{\DNSRateUnit}{\,\mathrm{Gpc}^{-3}\,\mathrm{yr}^{-1}}

\newcommand{\DNSChunkOneVT}{\ensuremath{0.00052}}
\newcommand{\DNSChunkOneVTLow}{\ensuremath{0.000085}}
\newcommand{\DNSChunkOneVTHigh}{\ensuremath{0.00012}}
\newcommand{\DNSvsAllSkyVTRatio}{\ensuremath{1.59}}

\newcommand{\DNSFirstHalfFourthObservationVT}{\ensuremath{0.0126}}
\newcommand{\DNSFirstHalfFourthObservationVTLow}{\ensuremath{0.0021}}
\newcommand{\DNSFirstHalfFourthObservationVTHigh}{\ensuremath{0.0030}}

\newcommand{\DNSVTFalseAlarmThreshold}{\ensuremath{1\;\mathrm{per~year}}}
\newcommand{\DNSVTpAstroThreshold}{\ensuremath{0.5}}
\newcommand{\AccumulatedVT}{\ensuremath{0.02056~\,\mathrm{Gpc}^{3}\,\mathrm{yr}^{1}}}
\newcommand{\AccumulatedVTError}{\ensuremath{0.00300~\,\mathrm{Gpc}^{3}\,\mathrm{yr}^{1}}}
\newcommand{\GWTCOneTwoThreeVT}{\ensuremath{0.00798~\,\mathrm{Gpc}^{3}\,\mathrm{yr}^{1}}}
\newcommand{\GWTCOneTwoThreeVTError}{\ensuremath{0.000178~\,\mathrm{Gpc}^{3}\,\mathrm{yr}^{1}}}

\newcommand{\DNSNTwoRateFifthPercentile}{\ensuremath 78}
\newcommand{\DNSNTwoRateNintyFifthPercentile}{\ensuremath 163}
\newcommand{\DNSNTwoRateMedian}{\ensuremath 106}

\newcommand{\DNSNTwoRateLowerBound}{\ensuremath 28}
\newcommand{\DNSNTwoRateUpperBound}{\ensuremath 269}

\newcommand{\DNSNThreeRateFifthPercentile}{\ensuremath 101}
\newcommand{\DNSNThreeRateNintyFifthPercentile}{\ensuremath 188}
\newcommand{\DNSNThreeRateMedian}{\ensuremath 154}

\newcommand{\DNSNThreeRateLowerBound}{\ensuremath 53}
\newcommand{\DNSNThreeRateUpperBound}{\ensuremath 342}

\begin{abstract}
We present a sub-threshold search for gravitational-wave inspirals from binary neutron stars using data from the first part of the fourth LIGO--Virgo--KAGRA observing run. 
To enhance sensitivity to this targeted population, we adopt a population prior informed by the Galactic double neutron star population observed via radio pulsar surveys and by the confirmed GW170817 event. 
The search, with a sensitivity $\sim 60\%$ higher than standard GWTC-4.0 searches, recovers the previously sub-threshold candidate GW231109\_235456 (reported in GWTC-4.0) with a network signal-to-noise ratio of \NetworkSNRSigTrigger\ and an improved false-alarm rate of 1 per 60 years. Accounting for a trials factor of five, arising from the four previous searches and this new search, the false-alarm rate of the candidate is approximately 1 per 12 years. 
If this event is of astrophysical origin, the inferred source properties indicate component masses of \HSMassoneSourceZeroPercentile $~M_\odot$ to \HSMassoneSourceNinetyPercentile $~M_\odot$ for the primary and \HSMasstwoSourceTenPercentile $~M_\odot$ to \HSMasstwoSourceHundredPercentile $~M_\odot$ for the secondary, yielding a total mass of $\HSTotalMassSourceFiftyPercentile^{+\HSTotalMassSourceNinetyFivePercentile}_{-\HSTotalMassSourceFivePercentile}~M_\odot$.
The event was localized to a region of \HSLocalizationAreaNinteyPercent$~\mathrm{deg}^2$ (90\% probability) at a luminosity distance of $\HSLuminosityDistanceFiftyPercentile^{+\HSLuminosityDistanceNinetyFivePercentile}_{-\HSLuminosityDistanceFivePercentile}~\mathrm{Mpc}$.
Assuming the signal arises from a binary neutron star merger, this event would contribute a local merger rate of $\DNSNThreeRateLowerBound~\DNSRateUnit$ to $\DNSNThreeRateUpperBound~\DNSRateUnit$ (90\% credible interval).
\end{abstract}


\section{Introduction}
\label{sec:intro}
The first \ac{BNS} merger, GW170817~\citep{LIGOScientific:2017vwq}, was discovered by \ac{LIGO}~\citep{LIGOScientific:2014pky} and Virgo~\citep{VIRGO:2014yos} more than nine years ago. 
As the first multi-messenger event, it was observed in coincidence with GRB170817A~\citep{LIGOScientific:2017vwq} and subsequently followed across the electromagnetic spectrum~\citep{Pian:2017gtc, Drout:2017ijr, LIGOScientific:2017adf, LIGOScientific:2017ync}. 
A second \ac{BNS} candidate, GW190425~\citep{LIGOScientific:2020aai}, was detected more than seven years ago, though no electromagnetic counterpart was identified. 
Since then, no significant \ac{BNS} candidates have been reported in real-time searches or the latest \ac{GWTC} releases~\citep{LIGOScientific:2026ifv, LIGOScientific:2026wfs}, despite the unprecedented detector sensitivity achieved during LIGO-Virgo-KAGRA \ac{O4}~\citep{LIGO:2024kkz, Capote:2024rmo, gwtc4-intro}. 
This apparent scarcity motivates searches for weaker signals in the existing data. 
One may ask: ``Could there be fainter \ac{BNS} signals hidden in our data?''

To answer this question, we incorporate a population-informed model for
\ac{BNS} systems as prior into our \ac{GW} search design. Previous analyses have
adopted broad compact-binary priors spanning both neutron star and black
hole systems~\citep{gwtc4-methods}, potentially trading sensitivity to
\ac{BNS} signals for broader coverage across multiple source classes,
rather than optimizing for the observed \ac{BNS} population~\citep{KAGRA:2021duu,
Ozel:2016oaf, Landry:2021hvl, Farrow:2019xnc, Alsing:2017bbc}.

To construct this population model, we consider data from the radio-observed Galactic \ac{BNS} systems together with the \ac{GW} merger event GW170817, which is consistent with the Galactic population~\citep{LIGOScientific:2017vwq, LIGOScientific:2018hze, Zhang:2019bhn, Galaudage:2020zst}. 
We model our target population with a bimodal Gaussian distribution, informed by the most recent population analysis of radio data~\citep{Farrow:2019xnc}, and numerically fit the observed component masses of the selected events to this model. 
The details of the population model are described in Sec.~\ref{sec:population}.

Meanwhile, GW190425 has component masses individually consistent with the Galactic \ac{BNS} population, but a total mass that lies outside that distribution at the $5\sigma$ level~\citep{LIGOScientific:2020aai}. 
This discrepancy has motivated interpretations invoking either a fast-merging neutron star population or the presence of one or more black-hole components~\citep{Romero-Shaw:2020aaj, Safarzadeh:2020efa, Foley:2020kus, Kyutoku:2020xka, Khadkikar:2025awf}.  
Given GW190425's unknown formation history, existing models struggle to reconcile it with the Galactic \ac{BNS} population within a single framework~\citep{Landry:2021hvl, Foley:2020kus}. 
To assess its influence on our result, App.~\ref{sec:other_pop_models} presents a refit of the bimodal Gaussian distribution that also includes the component masses of GW190425, assuming a \ac{BNS} origin.

We incorporate this model as a prior in the ranking statistic of a \ac{GW} search, following the source-dependent framework of \citet{Fong2018} (App.~\ref{sec:mass_model_math}). 
A similar sub-threshold search during the \ac{O1} yielded no candidates~\citep{Magee:2019vmb}; our approach offers two key improvements over that search. 
First, our bimodal Gaussian better reflects the observed Galactic \ac{BNS} mass distribution than the narrow Gaussian used previously, and includes GW170817 in the fit. 
Second, we account for cosmological redshift by sampling a uniform merger rate in comoving volume, rather than shifting the distribution's mean as in \citet{Magee:2019vmb}.

Applying our search to the LIGO-Virgo-KAGRA \ac{O4a} data~\citep{LIGOScientific:2025snk}, we identify a significant \ac{BNS} candidate, GW231109\_235456. 
We further perform parameter estimation using an agnostic mass prior to characterize this candidate, and use it to refine \ac{BNS} merger rate constraints.

\section{Search Design}
\label{sec:search_design}
We search for \ac{GW} signals from potential Galactic-like \ac{BNS} mergers using the \texttt{GstLAL} matched-filtering pipeline~\citep{Cannon:2011rj, Messick:2016aqy, Sachdev:2019vvd, Hanna:2019ezx, Cannon:2020qnf, Sakon:2022ibh, Tsukada:2023edh, Ewing:2023qqe}, which correlates \ac{GW} detector data with banks of waveform templates~\citep{Owen:1995tm, Hanna:2023}. 
This pipeline, which produced the results in \ac{GWTC}-4.0~\citep{gwtc4}, is used here with its latest improvements~\citep{Joshi:2025zdu, Joshi:2025nty, gwtc4-methods}.
Our analysis differs from the standard \ac{GWTC}-4.0 analyses~\citep{gwtc4-methods} in two key respects: (i) we incorporate a Galactic \ac{BNS} population prior (Sec.~\ref{sec:population}) to enhance sensitivity to these systems, and (ii) we use a denser template bank within the Galactic \ac{BNS} mass range (Sec.~\ref{sec:param_space}) to reduce signal loss.

Candidate triggers are ranked using the \texttt{GstLAL} likelihood-ratio statistic~\citep{Cannon:2012zt, Tsukada:2023edh}, which incorporates our prior choice. 
App.~\ref{sec:mass_model_math} illustrates how our prior enters as a source-dependent term for the \texttt{GstLAL} ranking statistic through the framework of \citet{Fong2018}. 
This study focuses on specifying the \ac{BNS} mass distribution used to construct the $P(\vec{t_j}|H_s, \rho)$ term in~\eqref{eq:mass_model}.
We adopt a bimodal Gaussian for the component masses, whereas the \ac{GWTC}-4.0 analyses used a power-law distribution for the primary mass and a uniform distribution for the secondary mass~\citep{gwtc4-methods}. 
All other analysis settings and ranking-statistic configurations follow the \texttt{GstLAL} setup used in \ac{GWTC}-4.0~\citep{gwtc4-methods}.
\subsection{Galactic BNS Population}
\label{sec:population}
Motivated by our template-bank design (Sec.~\ref{sec:param_space}), we describe the fiducial \ac{BNS} population using the coordinates $\vec{\gamma} = \{m_1, m_2, \chi_\text{eff}\}$ when constructing the probability density function in~\eqref{eq:population_term}:
\begin{equation}
p(\vec{\gamma}) = p(m_1, m_2) \, p(\chi_{\text{eff}}| m_1, m_2).
\label{eq:p_gamma}
\end{equation}

\begin{figure}
       \centering
       \includegraphics[width=1.0\linewidth]{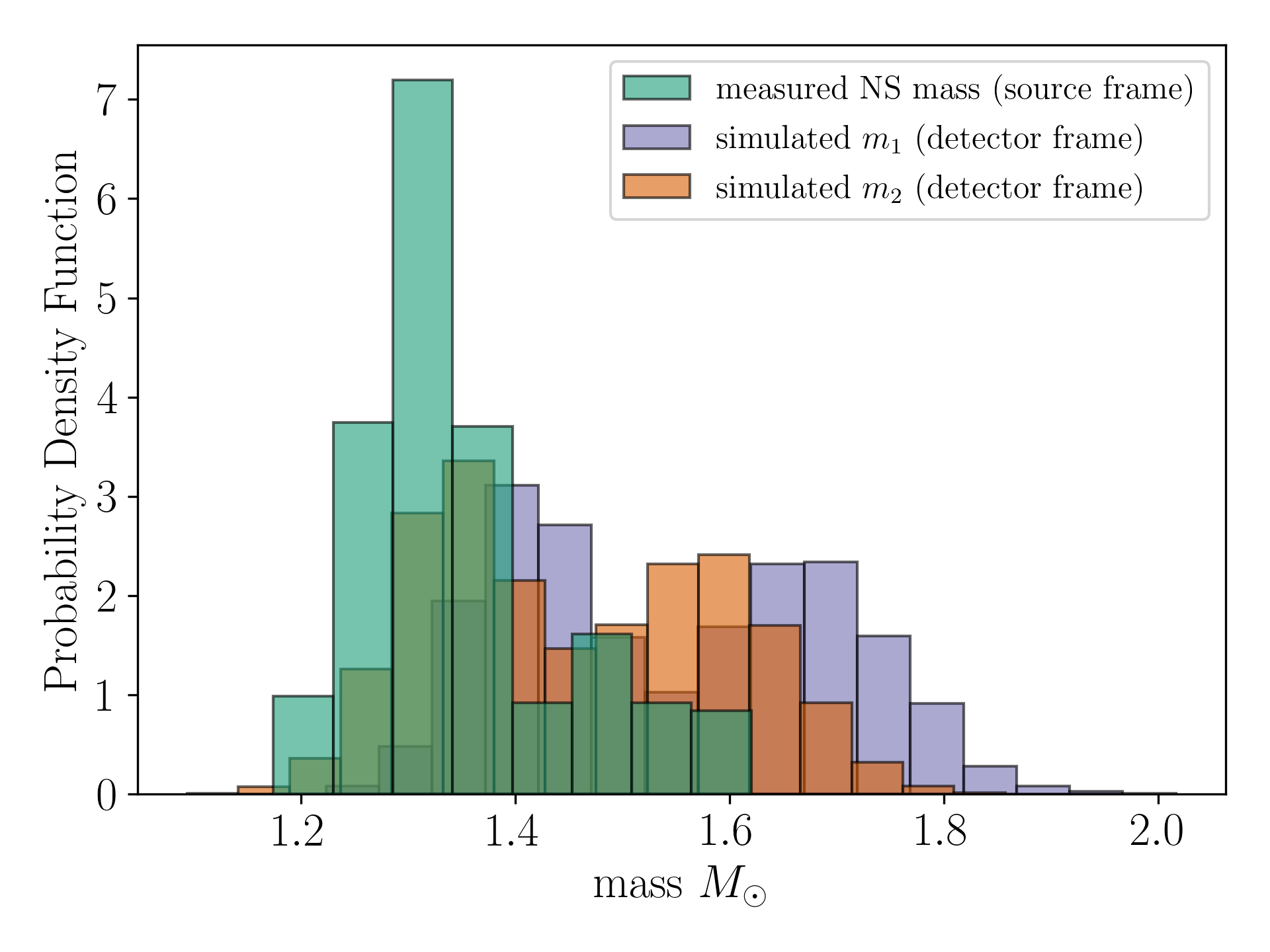}
       \caption{\textbf{Mass distribution of the Galactic \ac{BNS} population model adopted in this study.} 
       The green histogram shows the mass distribution from direct measurements of 13 Galactic \ac{BNS} systems and inferred masses of GW170817.
       The orange and purple histograms represent the simulated primary and secondary masses, respectively, sampled from the source-frame distribution truncated to $[1M_\odot, 2.17M_\odot]$ and mapped into the comoving volume.}
        \label{fig:dns_population}
\end{figure}

First, we seek a suitable mass distribution $p(m)$ to describe the neutron stars in radio-observed \ac{BNS} systems, one that also accommodates GW170817. 
Early studies found this population well described by a narrow Gaussian distribution~\citep{Thorsett:1998uc, Ozel:2012ax}, later broadened with the discovery of heavier systems~\citep{Kiziltan:2013ct, Ozel:2016oaf}. 
Following the observation of GW170817, \citet{Farrow:2019xnc} revisited this population and proposed a bimodal Gaussian model for the recycled neutron star in these binaries. 
While studies of \ac{GW} binaries containing at least one neutron star suggest a broader mass distribution with an excess of high-mass components~\citep{KAGRA:2021duu, Landry:2021hvl}, here we focus on the model consistent with radio observations.
As of May 2025, 19 Galactic \ac{BNS} systems have been identified in radio surveys, 13 of which have precisely measured component masses. 
Table~\ref{table:dns-catalog} summarizes these masses, together with the \ac{GW}-inferred masses of GW170817 and GW190425.
Figure~\ref{fig:dns_population} shows the distribution of measured component masses.

In this study, we do not distinguish between recycled and slow pulsars in Galactic \ac{BNS} systems, since such a distinction cannot be made in \ac{GW} searches. 
Given these conditions and assumptions, we infer the mass distribution by fitting the component mass measurements from the radio catalog and GW170817 to a bimodal Gaussian function:
\begin{equation}
\begin{split}
p(m) = w \cdot \frac{1}{\sqrt{2\pi \sigma_1^2}} \exp\left( -\frac{(m - \mu_1)^2}{2\sigma_1^2} \right) \\ 
+ (1 - w) \cdot \frac{1}{\sqrt{2\pi \sigma_2^2}} \exp\left( -\frac{(m - \mu_2)^2}{2\sigma_2^2} \right),
\end{split}
\label{eq:two_component_gaussian}
\end{equation}
where $w$ is the mixing weight for the first Gaussian component, $\mu_1$ and $\mu_2$ are the peaks of the two Gaussian components, and $\sigma_1$ and $\sigma_2$ are the widths of the two Gaussian components.
The fitting result gives $w=0.590$, $\mu_1=1.314\,M_\odot$, $\mu_2=1.40\,M_\odot$, $\sigma_1=0.0476\,M_\odot$, and $\sigma_2=0.123\,M_\odot$. 
Our model resembles that of \cite{Alsing:2017bbc}, although their study includes neutron star–white dwarf systems beyond Galactic \ac{BNS}, whose mergers lie outside the frequency range of LIGO-Virgo-KAGRA detectors \citep{Nelemans:2003xp}.

Next, we construct the mass distribution of neutron stars in binary systems, $p(m_1, m_2)$. 
We draw the component masses of each binary jointly from a bimodal Gaussian distribution, 
\begin{equation}
\begin{split}
p (m_1, m_2) \propto \frac{w}{2\pi\sigma_1^2}~\exp{(-\frac{(m_1-\mu_1)^2 + (m_2 - \mu_1)^2}{2\sigma_1^2})} \\
+ \frac{(1-w)}{2\pi\sigma_2^2}~\exp{(-\frac{(m_1-\mu_2)^2 + (m_2 - \mu_2)^2}{2\sigma_2^2})}, 
\end{split}
\label{eq:joint_two_component_gaussian}
\end{equation}
where the normalization of this distribution is constrained by two conditions:
\begin{itemize}
    \item $0.5 \leq m_2/m_1 \leq 1.0$
    \item $m_\text{min} \leq m_i \leq m_\text{max}$ where $m_\text{min}=1M_\odot$ and $m_\text{max}=2.17M_\odot$~\citep{Margalit:2017dij}
\end{itemize}

We sample the binary system using a joint model, rather than randomly pairing component masses drawn from the same distribution, to capture the correlation between the two neutron stars in the Galactic \ac{BNS} population.
Studies of Galactic \ac{BNS} and GW170817 formation channels have largely favored isolated binary evolution through a common-envelope phase~\citep{Tauris2017, Nair:2025kqv, Nair:2026byv}, in which the two neutron stars share the same metallicity, orbital history, and mass-transfer episodes. 
We expect $m_1$ and $m_2$ to be correlated. 
Since birth order is not identifiable from \ac{GW} observations alone, we adopt a symmetric same-peak correlation as a tractable proxy, drawing both masses from the same peak of the bimodal Gaussian. 
We can further interpret the bimodal peaks as two subpopulations of the Galactic \ac{BNS} systems: an electron-capture channel associated with the lower-mass peak $\mu_1$, and an iron-core-collapse channel associated with the heavier peak $\mu_2$~\citep{Chattaraj:2025rig, Schwab2010}. 
Coincidentally, such a model design assigns higher prior probability to GW190425-like events, where both components lie on the heavier peak of the bimodal Gaussian.
We adopt the fitted parameter values from~\eqref{eq:two_component_gaussian}.

\ac{GW} studies, by contrast, often adopt random pairing, drawing $m_1$ and $m_2$ independently from the same distribution~\citep{gwtc4-pop}, to remain agnostic to the various possible formation channels of \ac{GW}-observed \ac{BNS} systems.
We also construct a random-pairing model and present the results in App.~\ref{sec:random_pairing}. 
This model incidentally produces cross-peak pairs, providing a proxy for birth-order-asymmetric formation scenarios.

Moreover, we account for a redshift correction to our mass distribution, since \acp{GW} are observed in the (redshifted) detector frame; this effect is minor for GW170817 ($z \lesssim$ 0.1) and remains small throughout the observable volume of the LIGO-Virgo-KAGRA detectors for \ac{BNS} systems.
We incorporate this effect by sampling simulated events from
\begin{equation}
p(z) \propto \mathcal{R}(z)\frac{dV_c}{dz}\frac{1}{1+z},
\label{eq:p_z}
\end{equation}
assuming a constant source-frame merger rate and uniformity in comoving volume \citep{Essick:2025zed, Ye:2021klk, KAGRA:2021duu, gwtc4-pop}.
The source mass distribution is sampled from~\eqref{eq:joint_two_component_gaussian}, and the comoving volume element $dV_c/dz$~\citep{Ye:2021klk, Essick:2025zed} is evaluated under the standard cosmology~\citep{Planck:2015fie}. 
We further impose a minimum network \ac{SNR} of 8.0 to select a high-\ac{SNR} population, with sensitivity based on the measured \ac{LIGO} \ac{O4a} performance \citep{gwtc4-intro}.
Sampling is performed using the \texttt{monte-carlo-vt} \citep{Essick2025a} and \texttt{gw-distributions} \citep{Essick2025b} software packages.
Figure~\ref{fig:dns_population} shows the resulting sampled primary and secondary detector-frame masses.

For the sampled component masses $m^{\mathrm{det}}_1$ and $m^{\mathrm{det}}_2$, we assume each is still described by a bimodal Gaussian and fit them separately to~\eqref{eq:two_component_gaussian}, obtaining $w=0.48, \mu_1=1.41\,M_\odot, \sigma_1=0.057\,M_\odot, \mu_2=1.67\,M_\odot, \sigma_2=0.085\,M_\odot$ for $m^{\mathrm{det}}_1$ and $w=0.49, \mu_1=1.58\,M_\odot, \sigma_1=0.078\,M_\odot, \mu_2=1.34\,M_\odot, \sigma_2=0.06\,M_\odot$ for $m^{\mathrm{det}}_2$.
We can then write the mass distribution $p(m_1, m_2)$ in~\eqref{eq:p_gamma} as $p(m_1, m_2)=p(m^{\mathrm{det}}_1)p(m^{\mathrm{det}}_2)$. 

Finally, we assume a uniform distribution for the spin parameter,
\begin{equation}
p(\chi_{\mathrm{eff}} \mid m_1, m_2)
= \frac{1}{\chi_{\mathrm{eff,max}} - \chi_{\mathrm{eff,min}}},
\end{equation}
with the spin boundaries $\chi_{\mathrm{eff,max}}$ and $\chi_{\mathrm{eff,min}}$ set by our template bank (Table~\ref{table:design_bank}).

The resulting population model $p(\vec{\gamma})$ is further weighted by other factors to construct $P(\vec{t_j}|H_s, \rho)$ (App.~\ref{sec:mass_model_math}), which is computed using the \texttt{gwsci-manifold} \citep{manifold, Hanna:2023} software. 
Figure~\ref{fig:dns_mass_model} presents the corresponding $P(\vec{t_j}|H_s, \rho)$ likelihood across the search space. 

\subsection{Search Parameter Space}
\label{sec:param_space}
We construct a template bank targeting the mass range of Galactic \ac{BNS} systems. 
To improve sensitivity, we increase the template duration and raise the minimum match~\citep{Owen:1995tm}, thereby reducing signal loss due to template placement.
The bank is generated using \texttt{gwsci-manifold}~\citep{manifold, Hanna:2023}, which places templates directly in the $(m_1, m_2, \chi_{\texttt{eff}})$ coordinate space. 
Table~\ref{table:design_bank} summarizes the parameter space of this bank.
The subscript ‘det’ denotes the detector-frame quantities.
\begin{table}
\begin{center}
	\begin{tabular}{| l | l |}
		\hline
		Parameter & BNS Search \\
		\hline
		\hline
		Primary Mass, $m_1^{\mathrm{det}}$ & $\in [1.14, 2.1]~M_\odot$  \\
		Secondary Mass, $m_2^{\mathrm{det}}$ & $\in [1.1, 1.8]~M_\odot$  \\
		Mass Ratio, $q=m_2/m_1$ & $\in [0.67, 1.0]$  \\
        Chirp Mass, $\mathcal{M}^{\mathrm{det}}$ & $\in [1.0, 1.57]~M_\odot$  \\
        Total Mass, $M_T^{\mathrm{det}}$ & $ \in [2.3, 3.6]~M_\odot$ \\
		Dimensionless Component Spin, $\chi_{\rm {i,z}}$ & $ \in [-0.05, 0.05]$ \\
        Effective Spin, $\chi_{\texttt{eff}}$ & $ \in [-0.05, 0.05]$ \\
		Lower Frequency cut-off & 15 Hz \\
		Higher Frequency cut-off & 1024 Hz \\
		Waveform approximant & \texttt{IMRPhenomD} \\
		Minimum match & 99 $\%$ \\ 
		\hline
		Total number of templates & 360600 \\
		\hline
	\end{tabular}
	\caption{\textbf{Parameter space of the \ac{BNS} template bank.} Templates are placed in the mass ranges $m^\mathrm{det}_1 \in [1.14, 2.1]~M_\odot$ and $m^\mathrm{det}_2 \in [1.1, 1.8]~M_\odot$, with a mass ratio $q = m_2/m_1 \in [0.67, 1.0]$. 
    Component spins are set according to the low-spin assumption for \ac{BNS} mergers.
    Waveforms are modeled using \texttt{IMRPhenomD} over the frequency range 15 Hz to 1024 Hz.
    The subscript ``det'' denotes detector-frame quantities.}
	\label{table:design_bank} 
\end{center}
\end{table}

The mass bounds of the template bank are set by the sampled redshifted mass distribution described in Sec.~\ref{sec:population}, covering roughly a $\pm 3\sigma$ range. 
To mitigate edge effects at the boundaries of the template bank and ensure robust signal recovery across the target population, we extend the mass bounds to $[1.1,2.1]~M_\odot$.
The chirp mass, $\mathcal{M} = (m_1 m_2)^{3/5}/(m_1 + m_2)^{1/5}$, is implicitly constrained by the observed component-mass range.
Based on the measured systems listed in Table~\ref{table:dns-catalog}, we further restrict the bank to near-equal mass ratios and physically motivated total masses for Galactic \ac{BNS} systems. 
With these mass restrictions combined, our final bank also encompasses both the component and total masses of the outlier GW190425.

Radio observations imply that most Galactic \ac{BNS} systems have low spins at the time of merger \citep{Ozel:2016oaf, LIGOScientific:2017vwq}.
Accordingly, we restrict the dimensionless component
spins, $\chi_{i} = c |\vec{S}_{i}|/(Gm^2_{i})$, to the range $[-0.05,0.05]$ in the template bank, consistent with previous searches \citep{Sakon:2022ibh}.
We further assume that the component spins are aligned or anti-aligned with the orbital angular momentum and restrict the effective spin parameter $\chi_{\texttt{eff}}=(m_1\,\chi_{1,z} + m_2 \, \chi_{2, z})/(m_1 + m_2)$ to the same range of $[-0.05,0.05]$.

\begin{figure}
        \centering
        \includegraphics[width=1.05\linewidth]{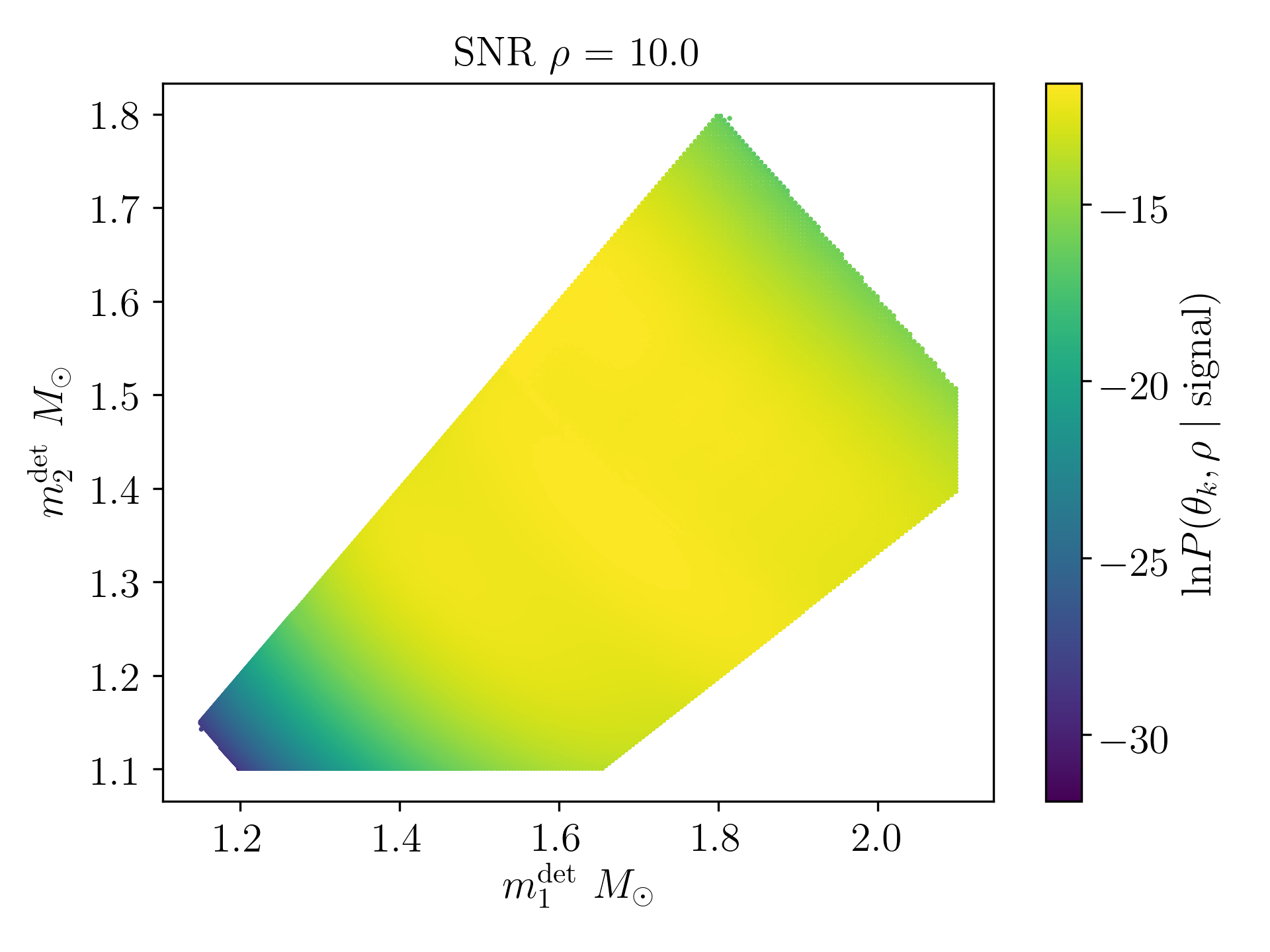}
        \caption{\textbf{Template bank designed for the \ac{BNS}-targeted search in the detector-frame $(m^\mathrm{det}_1, m^\mathrm{det}_2)$ parameter space.} 
        The color bar indicates the prior likelihood \citep{Fong2018} derived from the population model described in Sec.~\ref{sec:population}.}
        \label{fig:dns_mass_model}
\end{figure}

Compared with \ac{GWTC}-4.0 searches \citep{Sakon:2022ibh, gwtc4-methods}, we adopt a lower
frequency cutoff of $15~\mathrm{Hz}$ without imposing a waveform-duration
cutoff.
The upper frequency cutoff of the waveform is set to
$1024~\mathrm{Hz}$, and waveforms are modeled using \texttt{IMRPhenomD}
\citep{Khan:2016, Husa:2016}.
Additionally, we increase the minimum match used in template placement from 0.97 to 0.99, reducing the \ac{SNR} loss due to parameter discretization from $3\%$ to $1\%$.
Figure~\ref{fig:dns_mass_model} shows the resulting \ac{BNS} bank in the $(m^{\mathrm{det}}_1, m^{\mathrm{det}}_2)$ parameter space. 

\subsection{Search Sensitivity}
\label{sec:sensitivity}
We estimate the search sensitivity of our targeted analysis using the spacetime hypervolume, $\langle VT \rangle$, which quantifies the response of a search pipeline to a source population assumed to be uniformly distributed in comoving volume and source-frame time \citep{KAGRA:2021vkt, gwtc4-intro}.
Following standard practice, we generate simulated \ac{GW} signals, hereafter referred to as injections, and embed them into the search analysis.
For the \ac{BNS} sub-threshold search, we draw a population of binaries with component masses fixed at $(1.4 \, M_\odot, 1.4 \, M_\odot)$, isotopically distributed in luminosity distance $D_L$ between $10~\mathrm{Mpc}$ and $500~\mathrm{Mpc}$.
Because our search is sensitive only to signals in the local universe ($z \lesssim 0.1$) \citep{gwtc4-intro}, we neglect the effects of cosmological expansion in the sensitivity estimation.
To account for signal recoverability and the total observing time $T_{\mathrm{obs}}$ of our search, we estimate the spacetime hypervolume as
\begin{equation}
\langle VT \rangle \approx T_{\mathrm{obs}}~4\pi \int^{D_{\mathrm{L,max}}}_0 D_L^2~\epsilon(D_L)~dD_L
\label{eq:vt_equation_recovered}
\end{equation}
where $\epsilon(D_L) \equiv N_{\mathrm{found}}(D_L)/N_{\mathrm{inject}}(D_L)$ denotes the detection efficiency of the analysis at luminosity distance $D_L$.

Owing to computational resource constraints, the injection campaign was performed only for the first two weeks of our analysis, spanning \FirstHalfFourthObservationChunkOneStartTimeUTC~to \FirstHalfFourthObservationChunkOneEndTimeUTC.
This gives $\langle VT \rangle_{\text{two-week-BNS}}=\DNSChunkOneVT^{+\DNSChunkOneVTHigh}_{-\DNSChunkOneVTLow} \DNSVTUnit$ of our target search.
The uncertainty in this estimate is computed using the binomial proportion confidence interval.
An identical injection campaign was also performed for \texttt{GstLAL}’s \ac{GWTC}-4.0 search \citep{gwtc4-methods, gwtc4}, with the subscript “GWTC4” used to denote this case. 
Comparing the two results, we find that at a \ac{FAR} threshold of 1 per year,
\[
\frac{\langle VT \rangle_{\text{two-week-BNS}}}{\langle VT \rangle_{\text{two-week-GWTC4}}} = \DNSvsAllSkyVTRatio
\]
indicating an approximately 60\% sensitivity improvement of the target search for the \ac{BNS} population relative to the \ac{GWTC}-4.0 searches~\citep{gwtc4}.
To scale the two-week $\langle VT \rangle$ to the full \ac{O4a} observing period, we assume that the increase in $\langle VT \rangle$ grows linearly with the observing time $T$, so that
\[
\frac{\langle VT \rangle_{\text{two-week-BNS}}}{\langle VT \rangle_{\text{O4a-BNS}}}  = \frac{\langle VT \rangle_{\text{two-week-Network}}}{\langle VT \rangle_{\text{O4a-Network}}}
\]
where the subscript ``Network'' denotes the network-calibrated $\langle VT \rangle$ inferred from the \ac{BNS} horizon distance reported in \cite{gwtc4-intro}.
This gives $\langle VT \rangle_{\text{O4a-BNS}} \approx \DNSFirstHalfFourthObservationVT^{+\DNSFirstHalfFourthObservationVTHigh}_{-\DNSFirstHalfFourthObservationVTLow} \DNSVTUnit$ for our target search, with the associated uncertainty calculated using standard error propagation.
Compared with $\langle VT \rangle_{\text{O4a-GWTC4}}$ reported in \cite{gwtc4}, we find a sensitivity increase by a factor of $1.52$ at $(1.4\,M_\odot, 1.4 \, M_\odot)$, roughly consistent with the 60\% improvement observed in the two-week injection campaign.

\section{Results}
\label{sec:result}
We analyzed the full \ac{O4a} data from the \ac{LIGO} Hanford and Livingston Observatories, spanning \FirstHalfFourthObservationStartTimeUTC~to \FirstHalfFourthObservationEndTimeUTC. 
Data quality is described in~\citep{gwtc4-intro, gwtc4-methods}.
Adopting the population prior defined in Sec.~\ref{sec:population}, we identify one significant candidate in both detectors, with a \ac{FAR} of 1 per 60 years and a network \ac{SNR} of \NetworkSNRSigTrigger. 
This event appears in the low-latency GraceDB~\citep{gracedb} repository as the sub-threshold trigger S231109ci, and is reported as 
GW231109\_235456 in the Sec.~2.1.4 of \ac{GWTC}-4.0 sub-threshold catalog~\citep{gwtc4}.
Since \ac{GWTC}-4.0 presents the results from four independent searches for \ac{CBC} sources~\citep{gwtc4}, the inclusion of our targeted search increases the effective number of trials to five. 
Accounting for this trials factor, the effective \ac{FAR} for GW231109\_235456 becomes 1 per 12 years. 
App.~\ref{sec:far_justification} explains the \ac{FAR} discrepancy between our targeted search and \ac{GWTC}-4.0, clarifying why our search identifies this event significantly while \ac{GWTC}-4.0 searches do not. 
App.~\ref{sec:p_astro}, primarily Table~\ref{table:p_astro_table}, presents a binary p-astro classification for this event.

We further perform Bayesian parameter estimation using \texttt{Bilby}~\citep{bilby_paper, bilby_pipe_paper} to characterize the source, as described in Sec.~\ref{sec:result_pe}. 
Detection statistic and inferred source properties are summarized in Table~\ref{table:detection_candidate}. 
All reported \acp{CI} correspond to 90\%. 
Despite adopting an agnostic mass prior in parameter estimation, the inferred component masses are consistent with the range expected for Galactic \ac{BNS} systems.
Assuming a \ac{BNS} merger origin for GW231109\_235456, we estimate the corresponding merger rate using the sensitivity of our targeted search described in Sec.~\ref{sec:sensitivity}, with results presented in Sec.~\ref{sec:result_rates}. 

\begin{table}
\centering
\caption{\textbf{Properties of GW231109\_235456 identified in the \ac{O4a} sub-threshold search.} The table summarizes the detection statistics, event time, recovered template parameters from the search pipeline, and source properties inferred under low-spin and high-spin priors. For the source parameters, we report the (redshifted) detector-frame and source-frame chirp mass ($\mathcal{M}^\mathrm{det}$, $\mathcal{M}$), total mass ($M$), effective spin ($\chi_\mathrm{eff}$),  and luminosity distance ($D_L$) as medians with symmetric $90\%$ \acp{CI}; the primary mass ($m_1$), viewing angle ($\Theta$) and symmetric tidal deformability ($\Lambda_S$) as $0$--$90$th percentile \acp{CI}; the secondary mass ($m_2$) and mass ratio ($q$) as $10$–$100$th percentile \acp{CI}; and the localization area ($\Delta\Omega$) as a $90\%$ credible region.
The subscript ``det'' denotes detector-frame quantities.}
\label{table:detection_candidate}
\hspace*{-1.1cm}
\resizebox{10.0cm}{7.5cm}{%
\begin{tabular}{l r}
\toprule
\multicolumn{2}{c}{\textbf{Detection Statistics}} \\
\ac{FAR} (yr$^{-1}$) & \FalseAlarmRateSigTrigger \\
FAR with x5 trials factor (yr$^{-1}$) & \FalseAlarmRateGWTCFOURTrials \\
Hanford SNR & \HanfordSNRSigTrigger \\
Livingston SNR & \LivingstonSNRSigTrigger \\
\midrule
\multicolumn{2}{c}{\textbf{Time}} \\
UTC time & \UTCTimeSigTrigger \\
\midrule
\multicolumn{2}{c}{\textbf{Recovered Template Parameters}} \\
Detector-frame chirp mass, $\mathcal{M}^\mathrm{det}$ ($M_\odot$) & \ChirpMassSigTrigger \\
Detector-frame primary mass, $m_1^\mathrm{det}$ ($M_\odot$) & \PrimaryMassSigTrigger \\
Detector-frame secondary mass, $m_2^\mathrm{det}$ ($M_\odot$) & \SecondaryMassSigTrigger \\
\midrule
\multicolumn{2}{c}{\textbf{Inferred Parameter: Low-spin priors ($|\chi| \leq \LowSpinPriorUpperBound$)}} \\
Detector-frame chirp mass $\mathcal{M}^\mathrm{det}$ ($M_\odot$) & $\LSChirpMassFiftyPercentile^{+\LSChirpMassNinetyFivePercentile}_{-\LSChirpMassFivePercentile}$ \\
Chirp mass $\mathcal{M}$ ($M_\odot$) & $\LSChirpMassSourceFiftyPercentile^{+\LSChirpMassSourceNinetyFivePercentile}_{-\LSChirpMassSourceFivePercentile}$ \\
Primary mass $m_1$ ($M_\odot$) & \LSMassoneSourceZeroPercentile \,\text{ --}\, \LSMassoneSourceNinetyPercentile  \\
Secondary mass $m_2$ ($M_\odot$) & \LSMasstwoSourceTenPercentile \,\text{ --}\, \LSMasstwoSourceHundredPercentile \\
Mass ratio $q = m_2/m_1$ & \LSMassRatioTenPercentile \,\text{ --}\, \LSMassRatioHundredPercentile \\
Total mass $M$ ($M_\odot$) & $\LSTotalMassSourceFiftyPercentile^{+\LSTotalMassSourceNinetyFivePercentile}_{-\LSTotalMassSourceFivePercentile}$ \\
Effective spin $\chi_{eff}$ & $\LSChiEffFiftyPercentile^{+\LSChiEffNinetyFivePercentile}_{-\LSChiEffFivePercentile}$ \\
Luminosity distance $D_L$ (Mpc) & $\LSLuminosityDistanceFiftyPercentile^{+\LSLuminosityDistanceNinetyFivePercentile}_{-\LSLuminosityDistanceFivePercentile}$ \\
Viewing angle $\Theta$ (deg) & \LSViewingAngleZeroPercentile \,\text{ --}\, \LSViewingAngleNinetyPercentile \\
Localization area $\Delta\Omega$ (deg$^2$) & \LSLocalizationAreaNinteyPercent \\
Symmetric tidal deformability $\Lambda_S$ & \LSLambdaSymmetricZeroPercentile \,\text{ --}\, \LSLambdaSymmetricNinetyPercentile \\
\midrule
\multicolumn{2}{c}{\textbf{Inferred Parameter: High-spin priors ($|\chi| \leq \HighSpinPriorUpperBound$)}} \\
Detector-frame chirp mass $\mathcal{M}^\mathrm{det}$ ($M_\odot$) &  $\HSChirpMassFiftyPercentile^{+\HSChirpMassNinetyFivePercentile}_{-\HSChirpMassFivePercentile}$ \\
Chirp mass $\mathcal{M}$ ($M_\odot$) &  $\HSChirpMassSourceFiftyPercentile^{+\HSChirpMassSourceNinetyFivePercentile}_{-\HSChirpMassSourceFivePercentile}$ \\
Primary mass $m_1$ ($M_\odot$) &  \HSMassoneSourceZeroPercentile \,\text{ --}\, \HSMassoneSourceNinetyPercentile \\
Secondary mass $m_2$ ($M_\odot$) & \HSMasstwoSourceTenPercentile \,\text{ --}\, \HSMasstwoSourceHundredPercentile \\
Mass ratio $q = m_2/m_1$ &  \HSMassRatioTenPercentile \,\text{ --}\, \HSMassRatioHundredPercentile \\
Total mass $M$ ($M_\odot$) & $\HSTotalMassSourceFiftyPercentile^{+\HSTotalMassSourceNinetyFivePercentile}_{-\HSTotalMassSourceFivePercentile}$ \\
Effective spin $\chi_\mathrm{eff}$ & $\HSChiEffFiftyPercentile^{+\HSChiEffNinetyFivePercentile}_{-\HSChiEffFivePercentile}$\\
Luminosity distance $D_L$ (Mpc) & $\HSLuminosityDistanceFiftyPercentile^{+\HSLuminosityDistanceNinetyFivePercentile}_{-\HSLuminosityDistanceFivePercentile}$ \\
Viewing angle $\Theta$ (deg) & \HSViewingAngleZeroPercentile \,\text{ --}\, \HSViewingAngleNinetyPercentile \\
Localization area $\Delta\Omega$ (deg$^2$) & \HSLocalizationAreaNinteyPercent \\
Symmetric tidal deformability $\Lambda_S$ & \HSLambdaSymmetricZeroPercentile \,\text{ --}\, \HSLambdaSymmetricNinetyPercentile \\
\bottomrule
\end{tabular}
}
\end{table}

\subsection{Source Properties}
\label{sec:result_pe}
We perform source inference on 256 seconds of \ac{O4a} data containing GW231109\_235456 over the frequency range $20.0$ Hz to $972.8$ Hz.
Owing to the moderate \ac{SNR} of this event, extending the analysis to higher frequencies yields only a negligible increase in \ac{SNR}, consistent with the inferred parameter distribution reported below. 
The signal is modeled using the \texttt{IMRPhenomPv2\_NRTidalv2} waveform approximant~\citep{Hannam:2013oca, Khan:2015jqa, Dietrich:2019kaq, Colleoni:2023ple}, which incorporates both spin-induced orbital precession and tidal interactions between the neutron stars. 
\texttt{Bilby} evaluates the likelihood using reduced-order quadrature~\citep{Smith:2016qas, Morisaki:2023kuq}, with noise spectra estimated using \texttt{BayesWave}~\citep{Littenberg:2014oda, Gupta:2023jrn} and calibration uncertainties marginalized~\citep{gwtc4-methods}. 

We consider two isotropic spin priors: low-spin ($|\chi| \leq \LowSpinPriorUpperBound$) and high-spin ($|\chi| \leq \HighSpinPriorUpperBound$).
Unlike the search's population-informed mass prior (Sec.~\ref{sec:population}), we adopt an agnostic mass prior for parameter estimation, allowing an independent test of whether the recovered source properties are consistent with the Galactic \ac{BNS} population.
The mass prior is therefore taken to be uniform in the two-dimensional space of (redshifted) detector-frame component masses, $m^{\mathrm{det}}_i$, within a broad range of $0.3$–$5.0~M_\odot$.
See Section 5.5 of~\citet{gwtc4-methods} for a description of the prior choices adopted in the \ac{GWTC} analyses, which we largely follow here.
Informed by the search template, the (redshifted) detector-frame chirp mass, $\mathcal{M}^{\det}$, is restricted to $1.30$–$1.31~M_\odot$ and the mass ratio to $q \geq 1/8$. 
Detector-frame masses are converted to source-frame values using the inferred luminosity distance and a standard cosmology~\citep{Planck:2015fie, Astropy:2022ucr}.

\begin{figure}
\centering
   \includegraphics[width=1.0 \linewidth]{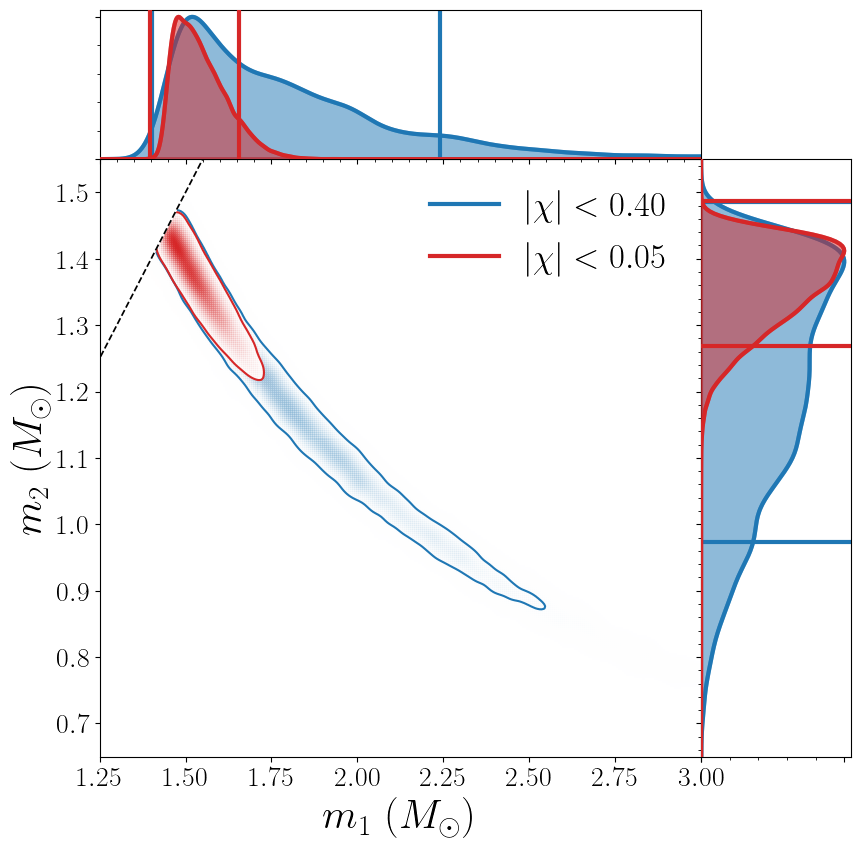}
   \caption{\textbf{Posterior distributions of the component masses, $m_1$ and $m_2$, for the high-spin ($|\chi| \leq 0.40$; blue) and low-spin ($|\chi| \leq 0.05$; red) source-inference analyses.}
   The dashed line in the two-dimensional panel denotes the equal-mass limit, $q = 1$.
   The vertical lines in the one-dimensional panels enclose $90\%$ of the posterior probability and correspond to the ranges reported in Table~\ref{table:detection_candidate}.
   The one-dimensional distributions are normalized to have equal maxima.
   }
    \label{fig:mass_posterior}
\end{figure}

Even with the agnostic mass prior, the inferred component masses (Fig.~\ref{fig:mass_posterior}) are consistent with the Galactic \ac{BNS} population~\citep{Ozel:2016oaf} and overlap with those of GW170817~\citep{LIGOScientific:2018hze}.
Under the high-spin prior, the posteriors remain consistent with the targeted population while also showing support extending beyond it.
For both priors, more prominently for the high-spin analysis, we infer a positive effective inspiral spin, $\chi_{\mathrm{eff}}$. 
Posterior summaries of source properties, sky localization, and viewing angle are given in Table~\ref{table:detection_candidate}.

None of the analyses presented here show posterior support near the prior boundaries (e.g. Fig.~\ref{fig:mass_posterior}), indicating that the results are driven by the data rather than by prior limits. 
Extending the priors to include masses comparable to GW190425 (in particular by allowing higher chirp masses) is therefore not expected to materially affect the inferred posterior for GW231109\_235456.

As a \ac{BNS}–targeted study, we account for tidal deformation by adopting a uniform prior on the symmetric combination $\Lambda_S = \tfrac{1}{2}(\Lambda_1 + \Lambda_2)$ over the range $0$–$5000$. 
We further impose a common, equation-of-state–independent relation between the individual tidal deformabilities $\Lambda_1$ and $\Lambda_2$~\citep{Yagi:2015pkc, Chatziioannou:2018vzf}, which are likewise restricted to $0$–$5000$.
The posterior of $\Lambda_S$ spans the full prior range but excludes values above \HSLambdaSymmetricNinetyPercentile \, (\LSLambdaSymmetricNinetyPercentile) at a 90\% \acp{CI} for the high-spin (low-spin) prior.
Additionally, this constraint can be recast as a bound on the combined dimensionless tidal deformability~\citep{LIGOScientific:2017vwq}:
\[
\tilde{\Lambda} = \frac{16}{13}\,
\frac{ (m_1 + 12 m_2)m_1^4 \Lambda_1 + (m_2 + 12 m_1)m_2^4 \Lambda_2 }
     { (m_1 + m_2)^5 }
\]
Re-weighting to a uniform prior in $\tilde{\Lambda}$ yields one-sided upper bounds of \HSLambdaTildeNinetyPercentileReweightBinaryLove \,(\LSLambdaTildeNinetyPercentileReweightBinaryLove).
Given this event's moderate \ac{SNR} and the absence of an electromagnetic counterpart, as the event was identified in a high-latency search, higher-order tidal parameters are less well constrained.

Given these factors, the tidal posteriors provide limited information on the neutron star equation of state from this event alone.
To assess this claim, we performed two additional analysis variants: one assuming both components are black holes ($\Lambda_1 = \Lambda_2 = 0$), and another allowing $\Lambda_1$ and $\Lambda_2$ to vary independently, in contrast to the main analysis above.
For the high-spin (low-spin) priors, the main analysis has a $\log_{10}$ Bayes factor of \HSlogTenBFBLoverBH \, (\LSlogTenBFBLoverBH)~relative to the black hole assumption and \HSlogTenBFBLoverfL \,  (\LSlogTenBFBLoverfL)~relative to the independent-$\Lambda$ assumption.

We also find a weak preference for the high-spin analysis over the low-spin one, with a $\log_{10}$ Bayes factor of \logTenBFHSoverLS.
We note that each of these comparisons involves models with somewhat different prior volumes: the black hole model omits tidal parameters entirely, the independent-$\Lambda$ model spans a different parameter space than the symmetric-$\Lambda_S$ parameterization used in the main analysis, and the high-spin and low-spin priors differ substantially in their allowed spin range. 
The reported Bayes factors should therefore be interpreted with these differences in mind.

In summary, for the moderately favored high-spin prior, we infer a weak preference for interpreting GW231109\_235456 as a \ac{BNS} system consistent with a common neutron star equation of state, relative to the two more general compact-object binary hypotheses.

\subsection{BNS Merger Rate}
\label{sec:result_rates}
Given the search result and the sensitivity estimated in Sec.~\ref{sec:sensitivity}, we can constrain the \ac{BNS} merger rate by treating GW231109\_235456 as a \ac{BNS} source.
The expected number of observed events $N$ is:
\begin{equation}
N = \mathcal{R} \langle VT \rangle,
\label{eq:expected_number}
\end{equation}
where the merger rate $\mathcal{R}$ is defined per unit comoving volume and per unit source-frame time~\citep{KAGRA:2021vkt}.
Observed events are modeled as a Poisson process with a Jeffreys prior, giving a posterior distribution for $\mathcal{R}$ that follows a Gamma distribution with shape parameter $N+0.5$ and scale parameter $1/\langle VT \rangle$. 
We estimate the merger rate under two scenarios: one including only GW170817 and GW231109\_235456 ($N=2$), and another including GW190425 as well ($N=3$). 
The accumulated sensitivity from all observing runs is:
\[
\langle VT \rangle = \langle VT \rangle_{\text{O1-O2-O3}} + \langle VT \rangle_{\text{O4a-BNS}}
\]
where $\langle VT \rangle_{\text{O1-O2-O3}} = \GWTCOneTwoThreeVT \pm \GWTCOneTwoThreeVTError$ represents the combined O1-O3 sensitivity~\citep{Essick:2025zed, LIGO:2025gwtc4vt}, mapped to a $(1.4\,M_\odot, 1.4\,M_\odot)$ \ac{BNS} system. 
The term $\langle VT \rangle_{\text{O4a-BNS}}$ denotes the sensitivity of our sub-threshold search (Sec.~\ref{sec:sensitivity}). 
The resulting accumulated sensitivity is~$\langle VT \rangle = \AccumulatedVT \pm \AccumulatedVTError$. 

We infer merger rates of $\mathcal{R}_{N=2} = \DNSNTwoRateMedian^{+\DNSNTwoRateNintyFifthPercentile}_{-\DNSNTwoRateFifthPercentile}~\mathrm{Gpc}^{-3}\,\mathrm{yr}^{-1}$ for the $N=2$ scenario and $\mathcal{R}_{N=3} = \DNSNThreeRateMedian^{+\DNSNThreeRateNintyFifthPercentile}_{-\DNSNThreeRateFifthPercentile}~\mathrm{Gpc}^{-3}\,\mathrm{yr}^{-1}$ for the $N=3$ scenario, with uncertainties corresponding to the 5th--95th percentile \acp{CI}.
These results are consistent with the \ac{BNS} merger rate from the \ac{GWTC}-3 population study~\citep{KAGRA:2021vkt, KAGRA:2021duu}, and statistically overlap with the rates reported in \ac{GWTC}-1~\citep{LIGOScientific:2018mvr}, \ac{GWTC}-2~\citep{LIGOScientific:2020kqk}, as well as the rates inferred for a GW170817-like population in the GW190425 discovery paper~\citep{LIGOScientific:2020aai}. 
They also overlap with the upper end of the \ac{BNS} merger rate range inferred under a null \ac{O4a}-detection scenario in the \ac{GWTC}-4.0 population analysis~\citep{gwtc4-pop}.

\section{Conclusion}
\label{sec:discussion}
In this work, we present a targeted search for \ac{GW} inspirals from the Galactic \ac{BNS} population including GW170817, using \ac{LIGO} \ac{O4a} data.
We identify a single candidate of interest, previously reported in the \ac{GWTC}-4.0 analyses as the sub-threshold event GW231109\_235456.
Within our search, noise is expected to produce a trigger more significant than GW231109\_235456 once every 60 years.
When considered alongside the four \ac{GWTC}-4.0 searches, and assuming independent noise realizations across searches, a trials factor of five reduces the significance to a \ac{FAR} of one per 12 years. 
This makes GW231109\_235456 the only candidate with \ac{FAR} below one per year in our analysis. 

We estimate that, for isotropically distributed \ac{BNS} systems, the sensitivity of this targeted search is approximately 60\% higher than that of the \ac{GWTC}-4.0 searches.
Using this sensitivity and the previously reported O1-O3 sensitivities for a ($1.4 \, M_\odot$, $1.4 \, M_\odot$) \ac{BNS} system, we infer the corresponding \ac{BNS} merger rate.
Although the low \ac{SNR} of GW231109\_235456 precludes a definitive identification, the inferred component masses and event rates are consistent with known \ac{BNS} systems. 
We encourage a systematic review of potential multi-messenger counterparts to this sub-threshold candidate, which could help clarify its physical nature.
Data supporting the findings of this study, including candidate triggers and parameter-estimation samples for GW231109\_235456, are publicly available at \textbf{https://zenodo.org/records/17645850}.

We adopt a Galactic-like \ac{BNS} prior tailored to this targeted search, whereas standard \ac{GWTC} searches, including low-latency analyses, consider a broad range of compact-binary priors.
This study highlights the sensitivity trade-offs among different subpopulations in a unified compact-binary search and demonstrates how targeted population assumptions can enhance sensitivity to specific source classes.
To improve sensitivity to the Galactic \ac{BNS} population while preserving overall search performance, we recommend the \texttt{GstLAL} search team explore the incorporation of our targeted prior into the standard analysis framework.
These developments may enhance searches during the upcoming \ac{O5} run and support future real-time, multi-messenger \ac{BNS} detections.

\section*{Acknowledgments}
We gratefully acknowledge Jocelyn Read for her insightful review of this paper.
This research has made use of data, software and/or web tools obtained from the
Gravitational Wave Open Science Center (https://www.gw-openscience.org/), a
service of \ac{LIGO} Laboratory, the \ac{LSC} and the Virgo
Collaboration.  
We especially made heavy use of the \ac{LVK} Algorithm
Library. 
\ac{LIGO} was constructed by the California Institute of Technology and the 
Massachusetts Institute of Technology with funding from the United States 
National Science Foundation (NSF) and operates under cooperative agreements 
PHYS-$0757058$ and PHY-$0823459$.
In addition, the Science and Technology Facilities Council (STFC) of the United Kingdom, the Max-Planck-Society (MPS), and the State of Niedersachsen/Germany supported the construction of \ac{aLIGO} and construction and operation of the GEO600 detector. 
Additional support for \ac{aLIGO} was provided by the Australian Research Council.  
Virgo is funded, through the European Gravitational Observatory (EGO), by the French Centre National de Recherche Scientifique (CNRS), the Italian Istituto Nazionale di Fisica Nucleare (INFN) and the Dutch Nikhef, with contributions by institutions from Belgium, Germany, Greece, Hungary, Ireland, Japan, Monaco, Poland, Portugal, Spain.

This material is based upon work supported by NSF's LIGO Laboratory which is a major facility fully funded by the National Science Foundation.
The authors are grateful for computational resources provided by the 
the \ac{LIGO} Lab culster at the \ac{LIGO} Laboratory and supported by 
PHY-$0757058$ and PHY$-0823459$, the Pennsylvania State University's Institute 
for Computational and Data Sciences gravitational-wave cluster, 
and supported by 
OAC-$2103662$, PHY-$2308881$, PHY-$2011865$, OAC-$2201445$, OAC-$2018299$, 
and PHY-$2207728$.  
CJH and LT acknowledges support from the Nevada Center for Astrophysics, from NASA Grant No. 80NSSC23M0104.
CJH also acknowledges support from the National Science Foundation through the Award No.~PHY-2409727.
CH Acknowledges generous support from the Eberly College of Science, the 
Department of Physics, the Institute for Gravitation and the Cosmos, the 
Institute for Computational and Data Sciences, and the Freed Early Career Professorship.
MWC acknowledges support from the National Science Foundation with grant numbers PHY-2308862 and PHY-2117997.
JDEC, PB, AB, CM acknowledges support from the National Science Foundation with grant number PHY-2513124.
BC acknowledges support from the NSF Graduate Research Fellowship Program under Grant No. DGE 21-46756.
DS acknowledges support from NSF Grant PHY-2020275 (Network for Neutrinos, Nuclear Astrophysics, and Symmetries (N3AS)).

We acknowledge the use of ChatGPT and Claude for proofreading text and mathematical content during the preparation of this manuscript.

\clearpage
\appendix
This appendix presents two alternative Galactic \ac{BNS} population priors (Sec.~\ref{sec:other_pop_models}), the statistical framework underlying our search results (Sec.~\ref{sec:mass_model_math}), a justification for the \ac{FAR} discrepancy between the sub-threshold and \ac{GWTC}-4.0 searches (Sec.~\ref{sec:far_justification}), a p-astro estimation for GW231109\_235456 (Sec.~\ref{sec:p_astro}), and the radio-observed catalog used to construct the population's mass distribution (Sec.~\ref{sec:app_dns_catalog}). 

\vspace{0.2in}
\section{Alternative Galactic \ac{BNS} Population Models}
\label{sec:other_pop_models}
We consider two additional mass distributions for the Galactic \ac{BNS} population in binary systems and examine how each affects the significance of our sole \ac{BNS} candidate in \ac{O4a}, GW231109\_235456. The first still draws binary systems from a joint bimodal Gaussian, but fits the component masses of GW190425 into the underlying mass distribution $p(m)$.
The second draws binary systems through random pairing, while retaining the fit to $p(m)$ that excludes GW190425.

\subsection{Joint-Pairing Model Incorporating GW190425}
\label{sec:join_pair_include_GW190425}
As stated in Sec.~\ref{sec:intro}, there is currently no unified framework describing a population that includes both the Galactic \ac{BNS} and GW190425, since the formation channel and host galaxy of GW190425 remain unknown.
\citet{Nair:2026byv} suggests that GW190425 could be explained by an extension of the standard Galactic \ac{BNS} population into the high-mass tail.
We therefore hypothesize that GW190425 can be described by the same bimodal Gaussian distribution $p(m)$ used for the Galactic \ac{BNS} population, and jointly fit its inferred component masses with those of the Galactic \ac{BNS} and GW170817 using the distribution in~\eqref{eq:two_component_gaussian}
This fit yields $w=0.786, \mu_1=1.318 \, M_\odot, \sigma_1=0.0625 \, M_\odot, \mu_2=1.544 \, M_\odot, \sigma_2=0.086 \, M_\odot$.

We then sample the binary systems using~\eqref{eq:joint_two_component_gaussian} as before, with the fitted parameter values given above.
We apply the same redshift correction, \ac{SNR} cutoff, and detector-frame component mass fitting as in Sec.~\ref{sec:population}.
Under this population prior, GW231109\_235456 remains the candidate of highest significance, with a \ac{FAR} of one per 52 years before correcting for the trials factor.

\subsection{Random-Pairing Model}
\label{sec:random_pairing}
In this case, we still consider the neutron star mass distribution result from Sec.~\ref{sec:population}, with parameter values fitted using the Galactic \ac{BNS} and GW170817.
Meanwhile, instead of drawing from the joint model, we draw the component masses $m_1$ and $m_2$ individually from $p(m)$, i.e.:
\begin{equation}
p(m_1, m_2)=p(m_1)p(m_2|m_1)~\text{with}~p(m_i)\propto \sum_{j=1, 2}w_j \, \mathcal{N}(m_i; \mu_j, \sigma_j),
\end{equation}
where $p(m_2|m_1)$ is subject to the normalization conditions $0.5 \leq m_2/m_1 \leq 1.0$ and $p(m_1, m_2)$ is subject to overall normalization condition $m_\text{min} \leq m_i \leq m_\text{max}$ with $m_\text{min}=1\, M_\odot$ and $m_\text{max}=2.17 \, M_\odot$ (as in~\eqref{eq:joint_two_component_gaussian}). 
 
As noted in Sec.~\ref{sec:population}, such pairing produces more cross-peak pairs for birth-order-asymmetric scenarios than the joint model does, although it does so by ignoring the correlation between $m_1$ and $m_2$ that a common formation channel would impose.

After sampling the binary pairs, we apply the same redshift correction, \ac{SNR} cutoff, and detector-frame component mass fitting as in Sec.~\ref{sec:population}.
The fitted parameters for the detector-frame component masses are
$w=0.55, \mu_1=1.34\,M_\odot, \sigma_1=0.042\,M_\odot, \mu_2=1.47\,M_\odot, \sigma_2=0.097\,M_\odot$ for $m^{\mathrm{det}}_1$ and $w=0.73, \mu_1=1.295\,M_\odot, \sigma_1=0.043\,M_\odot, \mu_2=1.305\,M_\odot, \sigma_2=0.10\,M_\odot$ for $m^{\mathrm{det}}_2$.
Under this population prior, GW231109\_235456 remains the candidate of highest significance, though its significance drops to one per 7 years before correcting for the trials factor.

However, one concern with such binary pairing is that it is likely to reduce support for the high-mass tail of the Galactic \ac{BNS} population. 
Our \ac{BNS} candidate, whose primary mass lies in the heavier tail of the population, is therefore down-ranked when pairing randomly. 
This model would similarly reduce support for GW190425-like events, whose component masses both lie in the heavier peak.

\begin{figure}
       \centering
       \includegraphics[width=1.0\linewidth]{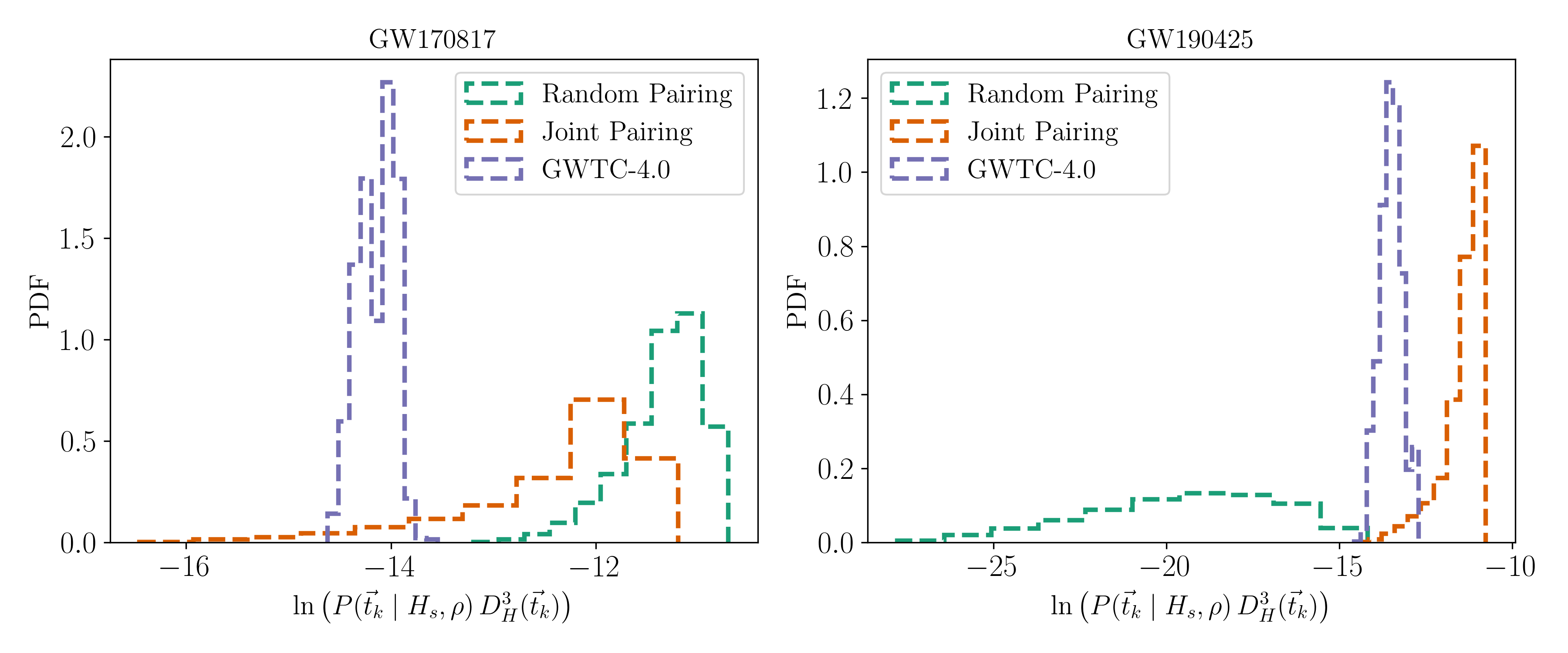}
       \caption{\textbf{The distribution of prior likelihoods weighted by observable volume, $P(\vec{t}_k \mid H_s, \rho)\, D_H^3(\vec{t}_k)$, evaluated at $\rho=10$, over the posterior mass ranges of GW170817 and GW190425.} 
       We compare three models: random pairing, joint pairing, and \ac{GWTC}-4.0. The joint-pairing model gives better support for both GW170817 and GW190425 than the \ac{GWTC}-4.0 prior, while the random-pairing model leaves GW190425 behind.}
        \label{fig:support_for_gw_events}
\end{figure}

We illustrate this in Fig.~\ref{fig:support_for_gw_events}, which plots the prior likelihood (\eqref{eq:mass_model}) weighted by the observable volume $D^3_H$ (i.e., $P(\vec{t}_k \mid H_s, \rho)\, D_H^3(\vec{t}_k)$), evaluated at $\rho=10$, over the posterior mass ranges of GW170817 and GW190425.
In the mass posterior of GW170817, both the random-pairing and joint-pairing models give more support overall (i.e., higher likelihood) than the \ac{GWTC}-4.0 prior.
Nevertheless, in the mass posterior of GW190425, the joint-pairing model still gives better support than the \ac{GWTC}-4.0 prior, while the random-pairing model falls well behind.
The search pipeline selects the template with the highest observed \ac{SNR} as the recovered template $\vec{t}_k$, so whether a similar signal is recoverable under these models still depends on the likelihood of the selected template.
But the distribution in the right panel of Fig.~\ref{fig:support_for_gw_events} suggests that support for GW190425-like events by random-pairing prior is likely to drop below that of both \ac{GWTC}-4.0 and this targeted search.

\subsection{Influence of Population-Prior Choice on the Significance of GW231109\_235456}
In this section, we assess how the significance of GW231109\_235456 varies with our assumptions about the population prior.
Comparing the three models presented in the main text and appendix, we find that whether we include the GW190425 posterior in our prior model has a moderate effect on the \ac{FAR} of GW231109\_235456.
We also find that the normalization over the allowed neutron-star mass range plays a negligible role.
We tested a prior model based purely on radio measurements, excluding both \ac{GW} events, using the joint-pairing scheme and a lower neutron-star mass cutoff, $m_\mathrm{max} = 1.8\, M_\odot$. 
This model gives a \ac{FAR} of one per 50 years for GW231109\_235456.
In contrast, the significance of GW231109\_235456 varies appreciably with our assumption about the pairing of component masses in the binary system.

How to pair the component masses for compact binary systems in \ac{GW} observations will depend on future discoveries of these events that reveal the population of component masses and the formation channels behind them.
In multi-source searches, we can adjust the relative prior volume through the observed event ratios; however, this is not possible for a targeted single-source search.

Even though the significance of GW231109\_235456 varies between the random-pairing and joint-pairing priors, their \acp{FAR} before the trials factor still remain below the one-per-year threshold typically used to flag \ac{BNS} candidates in \ac{GW} observations.
Moreover, the source properties obtained using an agnostic prior (Sec.~\ref{sec:result_pe}) show that this event is consistent with the known \ac{BNS} population, and we therefore recommend further follow-up studies on GW231109\_235456.

\section{Source-Dependent Ranking Statistic}
\label{sec:mass_model_math}
We briefly review the source-dependent ranking statistic~\citep{Fong2018} implemented in the \texttt{GstLAL} likelihood-ratio function; this source-dependence is what enables sub-threshold searches targeting a specific population, such as the Galactic \ac{BNS} population.
Derived from Bayesian formalism, the \texttt{GstLAL} likelihood-ratio function~\citep{Cannon:2015gha, Tsukada:2023edh} can be written as:
\[
\mathcal{L} = \dfrac{P\left(\vec{O}, \vec{\rho}, \vec{\xi}^2, \vec{t}, \vec{\phi}, \theta \mid \mathcal{H}_\mathrm{s}\right)}{P\left(\vec{O}, \vec{\rho}, \vec{\xi}^2, \vec{t}, \vec{\phi}, \theta \mid \mathcal{H}_\mathrm{n}\right)} = P(\vec{t_k} \mid \mathcal{H}_\mathrm{s}, \rho) \times \underbrace{\frac{P(...\mid \mathcal{H}_\mathrm{s})}{P(...\mid \mathcal{H}_\mathrm{n})}}_{\text{remaining terms}},
\]
where $P(\vec{t_k} \mid \mathcal{H}_\mathrm{s}, \rho)$ is considered as the source-dependent prior likelihood that models the probability of the trigger's intrinsic parameter $\theta$, approximated by its recovered (discretized) template $\vec{t_k}$, under the signal hypothesis $\mathcal{H}_\mathrm{s}$ and optimal \ac{SNR} $\rho$. 

This prior likelihood is constructed by combining the underlying source population model with corrections for noise migration and template-volume weighting due to discretization~\citep{Fong2018, Ray:2023nhx}, thereby accounting for both the intrinsic population distribution and observational selection effects.
\citet{Fong2018} first wrote out the formalism of $P(\vec{t_k} \mid \mathcal{H}_\mathrm{s}, \rho)$ as: 
\begin{equation}
P(\vec{t_k} \mid \mathcal{H}_\mathrm{s}, \rho) = \sum_j \left\{ P\left(\text{candidate in } \vec{t_k} \mid \mathcal{H}_\mathrm{s} = \rho\,\vec{t_j}\right) \times P(\vec{t_j} \mid \mathcal{H}_\mathrm{s})\right\}.
\label{eq:mass_model}
\end{equation}
Derivation and physical illustration of each term in~\eqref{eq:mass_model} can be found in Chapter 3 of~\citet{Fong2018}.

$P(\text{candidate in } \vec{t_k} \mid \mathcal{H}_\mathrm{s} = \rho\,\vec{t_j})$ in the equation is referred to as the noise migration factor, which describes the probability of the real signal $\vec{t_j}$ with optimal \ac{SNR} $\rho$ being recovered by template $\vec{t_k}$ under a Gaussian noise shift in the data~\citep{Fong2018}.
$P(\vec{t_j} \mid \mathcal{H}_\mathrm{s})$ is our prior assumption for the source population.
In this study, we adopt $P(\vec{t_j} \mid \mathcal{H}_\mathrm{s})$ for the Galactic \ac{BNS} population.
\citet{Fong2018} writes this term as
\begin{equation}
P(\vec{t_j}|\mathcal{H}_s) = \int_{V_{\vec{t_j}}} p(\vec{\gamma})d\vec{\gamma}
\end{equation}
and its discretization over template spaces as
\begin{equation}
P(\vec{t_j}|\mathcal{H}_s) = \sum_{V_i\in\mathcal{V}_j} p(\vec{\gamma_i})V_i
\label{eq:population_term}
\end{equation}
Here, $p(\vec{\gamma})$ is a fiducial function describing the distribution of our sources over the parameter space $\vec{\gamma}$.
For the \texttt{GstLAL} bank design, the parameter space is $\vec{\gamma} = \{m_1, m_2, \chi_\mathrm{eff}\}$.
Here, $i$ indexes the fiducial Voronoi point of the Voronoi cell $V_i$~\citep{Fong2018}.
$p(\vec{\gamma_i})$ is evaluated at the fiducial point of cell $i$.
$\mathcal{V}_j$ denotes the volume element occupied by template $j$~\citep{Fong2018}.
The prior likelihood is normalized over the entire bank space:
\[
\sum_k P(\vec{t_k}|H_s, \rho)=1.0
\]
Our population model in this work adopts a $p(\vec{\gamma})$ different from that of the \ac{GWTC}-4.0 prior (Sec.~\ref{sec:population}).
\section{\ac{FAR} Discrepancy of GW231109\_235456 between Targeted and \ac{GWTC}-4.0 Searches}
\label{sec:far_justification}
GW231109\_235456 is recovered by \texttt{GstLAL} in the standard \ac{GWTC}-4.0 searches with a \ac{FAR} of 631 per year~\citep{gwtc4}, while it is recovered much more significantly, with a \ac{FAR} of one per 60 years, in our targeted search.
How do we justify the \ac{FAR} discrepancy between the two search results?
Or, alternatively, why is GW231109\_235456 far more significant in our targeted search than in the \ac{GWTC}-4.0 searches?
We attribute this mainly to the difference in prior volume between the two searches over the \ac{BNS} mass range.

The \ac{FAR} ratio of GW231109\_235456 between our targeted search and \ac{GWTC}-4.0 search is:
\[
\frac{\text{FAR}_\mathrm{Targeted~Search}}{\text{FAR}_\mathrm{GWTC-4.0}} \approx 2.6\times10^{-5}
\]
To quantify the prior volume for the \ac{BNS} population, we sum the prior likelihood $P(\vec{t_k} \mid \mathcal{H}_\mathrm{s}, \rho)$ over \ac{BNS} templates, weighted by the observable volume $D_H^3(\vec{t_k})$:
\begin{equation}
    \sum_{\vec{t_k} \in \mathrm{BNS}} P(\vec{t_k} \mid \mathcal{H}_\mathrm{s}, \rho)\, D_H^3(\vec{t_k})
\end{equation}
Here, the \ac{BNS} mass range we adopt spans $m_\mathrm{min} = 1.0\, M_\odot$ to $m_\mathrm{max} = 2.17\, M_\odot$, slightly narrower than the range considered by \ac{GWTC} up to $m_\mathrm{max} = 3.0\, M_\odot$.
The prior volume ratio between our targeted search and \ac{GWTC}-4.0 search is:
\[
\frac{\sum_{\vec{t_k} \in \mathrm{BNS}} P(\vec{t_k} \mid \mathcal{H}_\mathrm{s} = \text{Galactic BNS}, \rho)\, D_H^3(\vec{t_k})}{\sum_{\vec{t_k} \in \mathrm{BNS}} P(\vec{t_k} \mid \mathcal{H}_\mathrm{s} = \text{GWTC-4.0}, \rho)\, D_H^3(\vec{t_k})} \approx 2 \times 10^{-5}
\]
which is of the same order of magnitude as the \ac{FAR} ratio.

Hence, we interpret our result as follows: since the \texttt{GstLAL} standard search in \ac{GWTC}-4.0 spreads its prior over the much broader parameter space of all compact-binary types (\ac{BNS}, \ac{NSBH}, and \ac{BBH}), its prior density within the \ac{BNS} mass range, including at the component masses of GW231109\_235456, is correspondingly diluted. 
Our targeted search, by concentrating its prior weight entirely over the narrower, astrophysically motivated Galactic \ac{BNS} mass range, assigns substantially more weight to this candidate's mass region, which explains its much higher significance. 
Both the resulting increase in sensitivity toward isotropically distributed \ac{BNS} sources (Sec.~\ref{sec:sensitivity}) and the complete lack of recoverability for \ac{NSBH} and \ac{BBH} sources in our targeted search reveal the sensitivity trade-off underlying this difference in prior density.

\section{P-Astro of GW231109\_235456}
\label{sec:p_astro}
In addition to the \ac{FAR}, \ac{GWTC} analyses also publish the significance of \ac{GW} candidates with their astrophysical probabilities, hereafter referred to as ``p-astro''~\citep{gwtc4, Farr:2013yna, Ray:2023nhx}. 
Readers of this work may also be interested in obtaining such an estimate for GW231109\_235456 based on the ranking-statistic result of our targeted search.
Since our targeted search is only sensitive to the \ac{BNS} population, we instead present a binary result, $p(\mathrm{BNS})$ vs.\ $p(\mathrm{noise})$, where
\[
p_{\mathrm{BNS}}(x) = \frac{\mathcal{R}_{\mathrm{BNS}}f(x)}{\mathcal{R}_{\mathrm{BNS}}f(x) + \mathcal{R}_{\mathrm{noise}}b(x)} = \frac{1}{1+ \dfrac{\mathcal{R}_{\mathrm{noise}}}{\mathcal{R}_{\mathrm{BNS}}}\dfrac{b(x)}{f(x)}}.
\]
Here, $\mathcal{R}_{\mathrm{noise}}$ and $\mathcal{R}_{\mathrm{BNS}}$ denote the noise and BNS rates of \ac{GW} candidates, respectively, while $b(x)/f(x)$ represents the ratio of the noise density function to the signal density function.
$x$ is the ranking-statistic parameter, for which we consider the likelihood ratio $\mathcal{L}$ used by \texttt{GstLAL}.

We adopt the FGMC p-astro formalism~\citep{Farr:2013yna}, following the same method as \ac{GWTC} analyses, which gives:
\begin{equation}
p_\mathrm{BNS}(x_k|\vec{x}_{N_r})=\frac{1}{1+\frac{P(x_k|\mathcal{H}_\mathrm{noise})}{P(x_k|\mathcal{H}_\mathrm{BNS})}\frac{\langle {R_\mathrm{noise}}\rangle_{N_r/k}}{\langle {R_\mathrm{BNS}}\rangle_{N_r/k}}}.
\label{eq:fgmc_pastro}
\end{equation}
$\frac{P(x_k|\mathcal{H}_\mathrm{noise})}{P(x_k|\mathcal{H}_\mathrm{BNS})}$ in~\eqref{eq:fgmc_pastro} is equivalent to $b(x)/f(x)$ for which we define:
\begin{equation}
f(\mathcal{L}) = \mathcal{L} \, b(\mathcal{L}) / A.
\end{equation}
We can estimate the $b(x)/f(x)$ through the normalization factor $A$:
\begin{equation}
1/A = \int_{\mathcal{L}^*}^\infty \mathcal{L} b(\mathcal{L}) d\mathcal{L}, 
\end{equation}
where $\mathcal{L}^*$ is the likelihood-ratio threshold we select and $b(\mathcal{L})$ is the noise distribution we obtain from the \texttt{GstLAL} ranking statistic of our targeted search. 
Therefore, 
\begin{equation}
\frac{b(\mathcal{L})}{f(\mathcal{L})} = \frac{A}{\mathcal{L}}
\end{equation}
These are the steps used to obtain the $b(x)/f(x)$ distribution for the \texttt{GstLAL} p-astro result of \ac{GWTC}-4.0 candidates, which we carefully follow in our calculation.

Next, we calculate the other quantities in~\eqref{eq:fgmc_pastro}. 
$x_k$ is the ranking-statistic data of the $k$-th trigger. 
$\vec{x}_{N_r}$ represents a collection of $N$ events in observing run $r$, where $\vec{x}_{N_r/k}$ represents the same collection of events excluding the $k$-th trigger.
Note that the $R$ notation here denotes the occurrence rate of events, not the merger rate $\mathcal{R}$ of the sources, in units of $\mathrm{yr}^{-1}$. 
\citet{Farr:2013yna} constructs the rate posteriors as
\begin{equation}
\langle {R_\beta}\rangle_{N_r/k}=\int d\vec{R} \, p(\vec{R}|\vec{x}_{N_r/k}) R_\beta
\label{eq:fgmc_rates_posterior}
\end{equation}
where $\beta$ denotes either source category or noise.
\texttt{GstLAL} p-astro analyses in \ac{GWTC} generally fix $\langle {R_\mathrm{noise}}\rangle=1$ for all triggers~\citep{Ray:2023nhx}. 
For binary classification, since we only have one source category, $p(\vec{R}|\vec{x}_{N_r/k})$ in~\eqref{eq:fgmc_rates_posterior} can be written as
\begin{equation}
p(R_\mathrm{BNS}|\vec{x}_{N_r/k}) \propto \underbrace{p(\vec{x}_{N_r/k} | R_\mathrm{BNS})}_{\text{likelihood}} \, \underbrace{p(R_\mathrm{BNS})}_{\text{prior}}
\end{equation}
The vector notation is dropped for a single source category.
In a \ac{CBC} source classification over the full population, we would instead consider $\vec{R} = \{\mathrm{BNS}, \mathrm{BBH}, \mathrm{NSBH}\}$, with the resulting integral evaluated via Gauss-Laguerre quadrature.
We construct the likelihood $p(\vec{x}_{N_r/k} | R_\mathrm{BNS})$ using the same FGMC method adopted by \ac{GWTC} analyses~\citep{Farr:2013yna}.

From the researcher's perspective, what really diverges between our analysis and the \ac{GWTC} p-astro calculation is the choice of the rate prior $p(R_\mathrm{BNS})$.
\ac{GWTC} analyses generally adopt uninformative priors, such as Uniform or Jeffreys~\citep{gwtc4-pop}.

Revisiting the design of our sub-threshold search and how it performs better than \ac{GWTC} searches for the \ac{BNS} population, we adopt a prior informed by the Galactic \ac{BNS} population. 
This informative prior is the major reason that our targeted search can ``dig'' harder for sub-threshold \ac{BNS} candidates than \ac{GWTC} searches, which is also reflected in the $VT$ improvement of our search toward \ac{BNS} (Sec.~\ref{sec:sensitivity}).
For our p-astro analysis, we therefore consider the \ac{BNS}-informed priors built from previously confirmed \ac{BNS} detections (GW170817 and GW190425). 
We do not include the \ac{O4a}-identified GW231109\_235456 in this prior construction, to reduce bias in our favor.
Table~\ref{table:p_astro_table} presents the $p_{\mathrm{BNS}}(x_k|\vec{x}_{N_r})$ of GW231109\_235456 from two uninformative priors and two \ac{BNS}-informed priors.

For the prior model, as in Sec.~\ref{sec:result_rates}, we consider a Gamma distribution, whose functional form matches the weight function of generalized Gauss-Laguerre quadrature, allowing the rate to be marginalized via 1D Gauss-Laguerre integration:
\begin{equation}
p(R_\mathrm{BNS}) = \Gamma \left(k=\alpha+1,\ s=\frac{VT_\mathrm{search}}{VT_\mathrm{prior}}\right) \propto R_\mathrm{BNS}^{\alpha}\,e^{-R_\mathrm{BNS}/s}
\end{equation}
where $k$ is the shape parameter, $s$ is the scale parameter, and $\alpha = \alpha_0 + N$.
$N$ is the number of previously confirmed \ac{BNS} detections used to inform the prior, independent of the search being analyzed.
For uninformative priors, we set $N=0$ (i.e., no external \ac{BNS} detections is incorporated). The Jeffreys prior chooses $\alpha_0=-0.5$, giving $\alpha=-0.5$, while the Uniform prior chooses $\alpha_0=0$, giving $\alpha=0$.
Note that in our rate estimation (Sec.~\ref{sec:result_rates}), we choose a shape of $k=N+0.5$, which is equivalent to $k=\alpha+1 = \alpha_0+N+1 = N+1-0.5 = N+0.5$ with the Jeffreys prior.
Uninformative priors incorporate no external detections, so there is no $VT_\mathrm{prior}$ to rescale against, and we set $s=1$.

For the prior informed by GW170817, we consider $N=1$ and adopt $\alpha_0=-0.5$ as a standard choice, giving $\alpha=0.5$ for the GW170817-informed prior.
For the prior informed by both GW170817 and GW190425, we consider $N=2$ with $\alpha_0=-0.5$, giving $\alpha=1.5$ for the GW170817- and GW190425-informed prior. 
For the scale $s$, we take the accumulated $\langle VT \rangle$ before our sub-threshold search $\langle VT\rangle_\mathrm{prior}= \langle VT\rangle_\mathrm{O1}+\langle VT\rangle_\mathrm{O2}+\langle VT\rangle_\mathrm{O3}$, mapped to the $(1.4\,M_\odot, 1.4\,M_\odot)$ mass pair, and $\langle VT\rangle_\mathrm{search}=\langle VT\rangle_\mathrm{O4a\text{-}BNS}$ as given in Sec.~\ref{sec:result_rates}:
\[
s = \frac{\langle VT \rangle_\mathrm{O4a\text{-}BNS}}{\langle VT \rangle_\mathrm{O1+O2+O3}} = 1.6
\]
Table~\ref{table:p_astro_table} reports the final calculated $p_{\mathrm{BNS}}$ for GW231109\_235456 using the priors and method described above.
The results from priors informed by GW170817 and GW190425 support a likely \ac{BNS} origin of GW231109\_235456 with probabilities higher than 90\%.
The uninformative priors, although less supportive than the \ac{BNS}-informed priors, still support a \ac{BNS} origin of this event with probabilities of $60.0$\% or higher.
\begin{table}[h]
\centering
\hspace*{-2.5cm}
\renewcommand{\arraystretch}{1.5}
\setlength{\tabcolsep}{15pt}
\begin{tabular}{|c|c|c|c|}
\hline
\textbf{Prior} & \textbf{Power $\alpha$} & \textbf{Scale $s$} & \textbf{$\boldsymbol{p}_{\textbf{BNS}}$(GW231109\_235456)} \\
\hline
Uniform & 0 & 1.0 & 74.70 \%\\
\hline
Jeffreys & -0.5 & 1.0 & 60.00\% \\
\hline
GW170817 & 0.5 & 1.6 & 90.26\% \\
\hline
GW170817 + GW190425 & 1.5 & 1.6 & 93.87\%\\
\hline
\end{tabular}
\caption{\textbf{$\boldsymbol{p}_{\textbf{BNS}}$ of GW231109\_235456 estimated using the FGMC formalism, with both uninformative and informative \ac{BNS} rate priors.} The binary p-astro results with priors informed by GW170817 and GW190425 support a \ac{BNS} origin of this event with a probability $\geq$ 90\%. The results with Uniform and Jeffreys priors support a \ac{BNS} origin of this event with probabilities of $60.0$\% or higher.} 
\label{table:p_astro_table}
\end{table}

\section{Galactic Double Neutron Star Catalog}
\label{sec:app_dns_catalog}
To construct the mass function of the Galactic \ac{BNS} population used in our ranking statistic (Sec.~\ref{sec:population}), we compiled the most recent double neutron star catalog available as of May 2025.
Table~\ref{table:dns-catalog} summarizes the total and component masses of these systems in the source frame.
Relative to previous studies, our catalog includes nine additional systems compared to \citet{Ozel:2016oaf} and three additional systems compared to \citet{Farrow:2019xnc}, after excluding one system likely to host a white-dwarf companion.
Mass measurements for these systems observed in radio surveys are derived from \ac{PK} parameters, which encode relativistic effects in pulsar binaries \citep{Ozel:2016oaf}.
The number of measurable \ac{PK} parameters depends on the pulsar timing precision and determines whether individual component masses can be inferred.
In our catalog, 13 systems have precise component-mass measurements, while six provide only the total mass, yielding a total of 19 events.

\begin{table}[!ht]
\begin{center}
    \hspace*{-1.75cm} 
	\begin{tabular}{| c | c | c | c | c |}
        \hline
        \multicolumn{5}{|c|}{\textbf{(a) Galactic Double Neutron Star systems with Precise Component Mass Measurement}} \\
        \hline
		\hline
		Pulsar Name & Total Mass ($M_\odot$) & Pulsar Mass ($M_\odot$) & Companion Mass ($M_\odot$) & Reference \\
		\hline
		J0453+1559 & 2.734 & 1.559 & 1.174 & \cite{Martinez:2015mya}  \\
		J0737-3039 & 2.58708 & 1.3381 & 1.2489 &  \cite{Kramer:2006nb} \\
		B1534+12 & 2.678463 & 1.3330 & 1.3455 &  \cite{Fonseca:2014qla} \\
		J1756-2251 & 2.56999 & 1.341 & 1.230 &  \cite{Ferdman:2014rna} \\
		J1906+0746 &  2.6134  & 1.291 & 1.322 &  \cite{vanLeeuwen:2014sca} \\
		B1913+16 & 2.828378 & 1.4398 & 1.3886 & \cite{Weisberg:2010zz} \\
		B2127 + 11c & 2.71279 & 1.358 & 1.354 &  \cite{Jacoby:2006dy} \\
        J1757-1854 & 2.73295 & 1.3384 & 1.3946 & \cite{Cameron:2017ody} \\
        J0509+3801 & 2.81 & 1.34 & 1.46 & \cite{Lynch:2018zxo} \\
        J1913+1102 & 2.8887 & 1.62 & 1.27 & \cite{Ferdman:2020huz} \\
        J1208-5936 & 2.586 & 1.26 & 1.32 & \cite{Bernadich:2023uru} \\
        J1518+4904 & 2.7186 & 1.470 & 1.248 & \cite{Tan:2024zvy} \\
        J1828+2456 & 2.59 & 1.306 & 1.299 & \cite{Haniewicz:2020jro} \\
        \hline
        \hline
        \multicolumn{5}{|c|}{\textbf{(b) Galactic Double Neutron Star Systems without Precise Component Mass Measurement}} \\
        \hline
        \hline
		Pulsar Name & Total Mass ($M_\odot$) & Pulsar Mass ($M_\odot$) & Companion Mass ($M_\odot$) &  Reference \\
		\hline
        J1811-1736 & 2.57 & $<$ 1.64 & $>$ 0.93 & \cite{Corongiu:2006rd} \\
        J1930-1852 & 2.59 & $<$ 1.32 & $>$ 1.30 & \cite{Swiggum:2015yra} \\
        J1946+2052 & 2.50 & $<$ 1.31 & $>$ 1.18 & \cite{Stovall:2018ouw} \\
        J1901+0658 & 2.79 & $<$ 1.68 & $>$ 1.11 & \cite{Su:2024gak} \\
        J1846-0513 & 2.6287 & $<$ 1.3455 & $>$ 1.2845 & \cite{Zhao:2024twu} \\
        J1411+2251 & 2.538 & $<$ 1.62 & $>$ 0.92 & \cite{Martinez:2017jbp} \\
        \hline
        \hline
        \multicolumn{5}{|c|}{\textbf{(c) \ac{BNS} Candidates of Gravitational Wave Events in GWTC}} \\
        \hline
        \hline
		Event Name & Total Mass ($M_\odot$) & Primary Mass ($M_\odot$) & Secondary Mass ($M_\odot$) & Reference \\
        \hline
        GW170817 (low-spin prior) & 2.74 & 1.36 - 1.60 & 1.17 - 1.36 & \cite{LIGOScientific:2018hze} \\
        GW170817 (high-spin prior) & 2.82 & 1.36 - 2.26 & 0.86 - 1.36 & \cite{LIGOScientific:2018hze} \\
        GW190425 (low-spin prior) & 3.3 & 1.60-1.87 & 1.46-1.69 & \cite{LIGOScientific:2020aai} \\
	GW190425 (high-spin prior) & 3.4 & 1.61-2.52 & 1.12-1.68 & \cite{LIGOScientific:2020aai} \\
	\hline
	\end{tabular}
	\caption{\textbf{Galactic double neutron star systems in source frame, including \ac{GW} \ac{BNS} candidate events, as of May 2025.}
    The catalog comprises a total of 19 systems identified through radio surveys.
    (a) catalog with precise component-mass measurements; this subset contains 13 systems.
    (b) catalog without precise component-mass measurements; this subset contains 6 systems, for which only the total mass is known due to insufficient determination of the \ac{PK} parameters.
    }
	\label{table:dns-catalog} 
\end{center}
\end{table}

\clearpage
\bibliographystyle{aasjournal}
\bibliography{references}

@article{Sakon:2022ibh,
    author = "Sakon, Shio and others",
    title = "{Template bank for compact binary mergers in the fourth observing run of Advanced LIGO, Advanced Virgo, and KAGRA}",
    eprint = "2211.16674",
    archivePrefix = "arXiv",
    primaryClass = "gr-qc",
    doi = "10.1103/PhysRevD.109.044066",
    journal = "Phys. Rev. D",
    volume = "109",
    number = "4",
    pages = "044066",
    year = "2024"
}

@misc{manifold,
	author = {Hanna, Chad},
	title = {manifold},
	year = {2024},
	publisher = {GitHub},
	journal = {GitHub repository},
	howpublished = {\url{https://git.ligo.org/chad-hanna/manifold}},
	commit = {dc651a26f0341317228b4918af2743c1162b8600}
}

@article{Hanna:2023,
	title = {Binary tree approach to template placement for searches for gravitational waves from compact binary mergers},
	author = {Hanna, Chad and others},
	journal = {Phys. Rev. D},
	volume = {108},
	issue = {4},
	pages = {042003},
	numpages = {7},
	year = {2023},
	month = {Aug},
	publisher = {American Physical Society},
	doi = {10.1103/PhysRevD.108.042003},
	url = {https://link.aps.org/doi/10.1103/PhysRevD.108.042003}
}

@article{Ewing:2023qqe,
    author = "Ewing, Becca and others",
    title = "{Performance of the low-latency GstLAL inspiral search towards LIGO, Virgo, and KAGRA{\textquoteright}s fourth observing run}",
    eprint = "2305.05625",
    archivePrefix = "arXiv",
    primaryClass = "gr-qc",
    doi = "10.1103/PhysRevD.109.042008",
    journal = "Phys. Rev. D",
    volume = "109",
    number = "4",
    pages = "042008",
    year = "2024"
}

@misc{Joshi:2025zdu,
    author = "Joshi, Prathamesh and others",
    title = "{How Many Times Should We Matched Filter Gravitational Wave Data? A Comparison of GstLAL's Online and Offline Performance}",
    eprint = "2505.23959",
    archivePrefix = "arXiv",
    primaryClass = "gr-qc",
    month = "5",
    year = "2025"
}

@misc{Joshi:2025nty,
    author = "Joshi, Prathamesh and others",
    title = "{New Methods for Offline GstLAL Analyses}",
    eprint = "2506.06497",
    archivePrefix = "arXiv",
    primaryClass = "gr-qc",
    month = "6",
    year = "2025"
}

@article{Messick:2016aqy,
    author = "Messick, Cody and others",
    title = "{Analysis Framework for the Prompt Discovery of Compact Binary Mergers in Gravitational-wave Data}",
    eprint = "1604.04324",
    archivePrefix = "arXiv",
    primaryClass = "astro-ph.IM",
    doi = "10.1103/PhysRevD.95.042001",
    journal = "Phys. Rev. D",
    volume = "95",
    number = "4",
    pages = "042001",
    year = "2017"
}

@article{Sachdev:2019vvd,
    author = "Sachdev, Surabhi and others",
    title = "{The GstLAL Search Analysis Methods for Compact Binary Mergers in Advanced LIGO's Second and Advanced Virgo's First Observing Runs}",
    eprint = "1901.08580",
    journal = "arXiv e-prints",
    archivePrefix = "arXiv",
    primaryClass = "gr-qc",
    month = "1",
    year = "2019"
}

@article{Cannon:2011rj,
    author = "Cannon, Kipp and Hanna, Chad and Keppel, Drew",
    editor = "Hannam, Mark and Sutton, Patrick and Hild, Stefan and van den Broeck, Chris",
    title = "{Interpolating compact binary waveforms using the singular value decomposition}",
    eprint = "1108.5618",
    archivePrefix = "arXiv",
    primaryClass = "gr-qc",
    reportNumber = "LIGO-P1100101-V2",
    doi = "10.1103/PhysRevD.85.081504",
    journal = "Phys. Rev. D",
    volume = "85",
    pages = "081504",
    year = "2012"
}

@article{Hanna:2019ezx,
    author = "Hanna, Chad and others",
    title = "{Fast evaluation of multidetector consistency for real-time gravitational wave searches}",
    eprint = "1901.02227",
    archivePrefix = "arXiv",
    primaryClass = "gr-qc",
    doi = "10.1103/PhysRevD.101.022003",
    journal = "Phys. Rev. D",
    volume = "101",
    number = "2",
    pages = "022003",
    year = "2020"
}

@article{Cannon:2020qnf,
    author = "Cannon, Kipp and others",
        title = "{GstLAL: A software framework for gravitational wave discovery}",
      journal = {SoftwareX},
         year = 2021,
        month = jun,
       volume = {14},
          eid = {100680},
        pages = {100680},
          doi = {10.1016/j.softx.2021.100680},
archivePrefix = {arXiv},
       eprint = {2010.05082},
 primaryClass = {astro-ph.IM},
}

@article{Ray:2023nhx,
    author = "Ray, Anarya and others",
    title = "{When to Point Your Telescopes: Gravitational Wave Trigger Classification for Real-Time Multi-Messenger Followup Observations}",
    eprint = "2306.07190",
    journal = "arXiv e-prints",
    archivePrefix = "arXiv",
    primaryClass = "gr-qc",
    reportNumber = "LIGO-P2300141",
    month = "6",
    year = "2023"
}

@article{Tsukada:2023edh,
    author = "Tsukada, Leo and others",
    title = "{Improved ranking statistics of the GstLAL inspiral search for compact binary coalescences}",
    eprint = "2305.06286",
    archivePrefix = "arXiv",
    primaryClass = "astro-ph.IM",
    doi = "10.1103/PhysRevD.108.043004",
    journal = "Phys. Rev. D",
    volume = "108",
    number = "4",
    pages = "043004",
    year = "2023"
}

@article{Magee:2019vmb,
    author = "Magee, Ryan and others",
    title = "{Sub-threshold Binary Neutron Star Search in Advanced LIGO{\textquoteright}s First Observing Run}",
    eprint = "1901.09884",
    archivePrefix = "arXiv",
    primaryClass = "gr-qc",
    doi = "10.3847/2041-8213/ab20cf",
    journal = "Astrophys. J. Lett.",
    volume = "878",
    number = "1",
    pages = "L17",
    year = "2019"
}

@article{LIGOScientific:2017vwq,
    author = "Abbott, B. P. and others",
    collaboration = "LIGO Scientific, Virgo",
    title = "{GW170817: Observation of Gravitational Waves from a Binary Neutron Star Inspiral}",
    eprint = "1710.05832",
    archivePrefix = "arXiv",
    primaryClass = "gr-qc",
    reportNumber = "LIGO-P170817",
    doi = "10.1103/PhysRevLett.119.161101",
    journal = "Phys. Rev. Lett.",
    volume = "119",
    number = "16",
    pages = "161101",
    year = "2017"
}

@article{LIGOScientific:2018hze,
    author = "Abbott, B. P. and others",
    collaboration = "LIGO Scientific, Virgo",
    title = "{Properties of the binary neutron star merger GW170817}",
    eprint = "1805.11579",
    archivePrefix = "arXiv",
    primaryClass = "gr-qc",
    doi = "10.1103/PhysRevX.9.011001",
    journal = "Phys. Rev. X",
    volume = "9",
    number = "1",
    pages = "011001",
    year = "2019"
}

@article{LIGOScientific:2020aai,
    author = "Abbott, B. P. and others",
    collaboration = "LIGO Scientific, Virgo",
    title = "{GW190425: Observation of a Compact Binary Coalescence with Total Mass $\sim 3.4 M_{\odot}$}",
    eprint = "2001.01761",
    archivePrefix = "arXiv",
    primaryClass = "astro-ph.HE",
    reportNumber = "LIGO-P190425",
    doi = "10.3847/2041-8213/ab75f5",
    journal = "Astrophys. J. Lett.",
    volume = "892",
    number = "1",
    pages = "L3",
    year = "2020"
}

@article{LIGOScientific:2014pky,
    author = "Aasi, J. and others",
    collaboration = "LIGO Scientific",
    title = "{Advanced LIGO}",
    eprint = "1411.4547",
    archivePrefix = "arXiv",
    primaryClass = "gr-qc",
    doi = "10.1088/0264-9381/32/7/074001",
    journal = "Class. Quant. Grav.",
    volume = "32",
    pages = "074001",
    year = "2015"
}

@article{VIRGO:2014yos,
    author = "Acernese, F. and others",
    collaboration = "VIRGO",
    title = "{Advanced Virgo: a second-generation interferometric gravitational wave detector}",
    eprint = "1408.3978",
    archivePrefix = "arXiv",
    primaryClass = "gr-qc",
    doi = "10.1088/0264-9381/32/2/024001",
    journal = "Class. Quant. Grav.",
    volume = "32",
    number = "2",
    pages = "024001",
    year = "2015"
}

@techreport{gracedb,
  author       = {LVK},
  title        = {GraceDB: The Gravitational-Wave Candidate Event Database},
  institution  = {LIGO Project},
  number       = {LIGO-T1400365-v5},
  year         = {2018},
  url          = {https://dcc.ligo.org/LIGO-T1400365/public},
}

@article{Galaudage:2020zst,
    author = "Galaudage, Shanika and Adamcewicz, Christian and Zhu, Xing-Jiang and Stevenson, Simon and Thrane, Eric",
    title = "{Heavy Double Neutron Stars: Birth, Midlife, and Death}",
    eprint = "2011.01495",
    archivePrefix = "arXiv",
    primaryClass = "astro-ph.HE",
    doi = "10.3847/2041-8213/abe7f6",
    journal = "Astrophys. J. Lett.",
    volume = "909",
    number = "2",
    pages = "L19",
    year = "2021"
}

@article{Foley:2020kus,
    author = "Foley, Ryan J. and Coulter, David A. and Kilpatrick, Charles D. and Piro, Anthony L. and Ramirez-Ruiz, Enrico and Schwab, Josiah",
    title = "{Updated Parameter Estimates for GW190425 Using Astrophysical Arguments and Implications for the Electromagnetic Counterpart}",
    eprint = "2002.00956",
    archivePrefix = "arXiv",
    primaryClass = "astro-ph.HE",
    doi = "10.1093/mnras/staa725",
    journal = "Mon. Not. Roy. Astron. Soc.",
    volume = "494",
    number = "1",
    pages = "190--198",
    year = "2020"
}

@article{Kyutoku:2020xka,
    author = "Kyutoku, Koutarou and Fujibayashi, Sho and Hayashi, Kota and Kawaguchi, Kyohei and Kiuchi, Kenta and Shibata, Masaru and Tanaka, Masaomi",
    title = "{On the Possibility of GW190425 Being a Black Hole{\textendash}Neutron Star Binary Merger}",
    eprint = "2001.04474",
    archivePrefix = "arXiv",
    primaryClass = "astro-ph.HE",
    doi = "10.3847/2041-8213/ab6e70",
    journal = "Astrophys. J. Lett.",
    volume = "890",
    number = "1",
    pages = "L4",
    year = "2020"
}

@article{Romero-Shaw:2020aaj,
    author = "Romero-Shaw, Isobel M. and Farrow, Nicholas and Stevenson, Simon and Thrane, Eric and Zhu, Xing-Jiang",
    title = "{On the origin of GW190425}",
    eprint = "2001.06492",
    archivePrefix = "arXiv",
    primaryClass = "astro-ph.HE",
    doi = "10.1093/mnrasl/slaa084",
    journal = "Mon. Not. Roy. Astron. Soc.",
    volume = "496",
    number = "1",
    pages = "L64--L69",
    year = "2020"
}

@article{Safarzadeh:2020efa,
    author = "Safarzadeh, Mohammadtaher and Ramirez-Ruiz, Enrico and Berger, Edo",
    title = "{Does GW190425 require an alternative formation pathway than a fast-merging channel?}",
    eprint = "2001.04502",
    archivePrefix = "arXiv",
    primaryClass = "astro-ph.HE",
    doi = "10.3847/1538-4357/aba596",
    journal = "Astrophys. J.",
    volume = "900",
    number = "1",
    pages = "13",
    year = "2020"
}

@article{Thorsett:1998uc,
    author = "Thorsett, S. E. and Chakrabarty, Deepto",
    title = "{Neutron star mass measurements. 1. Radio pulsars}",
    eprint = "astro-ph/9803260",
    archivePrefix = "arXiv",
    doi = "10.1086/306742",
    journal = "Astrophys. J.",
    volume = "512",
    pages = "288",
    year = "1999"
}

@article{Ozel:2012ax,
    author = "{\"O}zel, Feryal and Psaltis, Dimitrios and Narayan, Ramesh and Villarreal, Antonia Sierra",
    title = "{On the Mass Distribution and Birth Masses of Neutron Stars}",
    eprint = "1201.1006",
    archivePrefix = "arXiv",
    primaryClass = "astro-ph.HE",
    doi = "10.1088/0004-637X/757/1/55",
    journal = "Astrophys. J.",
    volume = "757",
    pages = "55",
    year = "2012"
}

@article{Kiziltan:2013ct,
    author = "Kiziltan, Bulent and Kottas, Athanasios and Thorsett, Stephen E.",
    title = "{The Neutron Star Mass Distribution}",
    eprint = "1011.4291",
    archivePrefix = "arXiv",
    primaryClass = "astro-ph.GA",
    doi = "10.1088/0004-637X/778/1/66", 
    journal = "The Astrophysical Journal",
    volume = "778", 
    month = "11",
    year = "2013"
}

@article{Ozel:2016oaf,
    author = {{\"O}zel, Feryal and Freire, Paulo},
    title = "{Masses, Radii, and the Equation of State of Neutron Stars}",
    eprint = "1603.02698",
    archivePrefix = "arXiv",
    primaryClass = "astro-ph.HE",
    doi = "10.1146/annurev-astro-081915-023322",
    journal = "Ann. Rev. Astron. Astrophys.",
    volume = "54",
    pages = "401--440",
    year = "2016"
}

@article{Farrow:2019xnc,
    author = "Farrow, Nicholas and Zhu, Xing-Jiang and Thrane, Eric",
    title = "{The mass distribution of Galactic double neutron stars}",
    eprint = "1902.03300",
    archivePrefix = "arXiv",
    primaryClass = "astro-ph.HE",
    doi = "10.3847/1538-4357/ab12e3",
    journal = "Astrophys. J.",
    volume = "876",
    number = "1",
    pages = "18",
    year = "2019"
}

@article{KAGRA:2021duu,
    author = "Abbott, R. and others",
    collaboration = "KAGRA, VIRGO, LIGO Scientific",
    title = "{Population of Merging Compact Binaries Inferred Using Gravitational Waves through GWTC-3}",
    eprint = "2111.03634",
    archivePrefix = "arXiv",
    primaryClass = "astro-ph.HE",
    reportNumber = "LIGO-P2100239 ; Data release: https://zenodo.org/record/5655785, LIGO-P2100239",
    doi = "10.1103/PhysRevX.13.011048",
    journal = "Phys. Rev. X",
    volume = "13",
    number = "1",
    pages = "011048",
    year = "2023"
}

@article{KAGRA:2021vkt,
    author = "Abbott, R. and others",
    collaboration = "KAGRA, VIRGO, LIGO Scientific",
    title = "{GWTC-3: Compact Binary Coalescences Observed by LIGO and Virgo during the Second Part of the Third Observing Run}",
    eprint = "2111.03606",
    archivePrefix = "arXiv",
    primaryClass = "gr-qc",
    reportNumber = "LIGO-P2000318",
    doi = "10.1103/PhysRevX.13.041039",
    journal = "Phys. Rev. X",
    volume = "13",
    number = "4",
    pages = "041039",
    year = "2023"
}

@article{LIGOScientific:2018mvr,
    author = "Abbott, B. P. and others",
    collaboration = "LIGO Scientific, Virgo",
    title = "{GWTC-1: A Gravitational-Wave Transient Catalog of Compact Binary Mergers Observed by LIGO and Virgo during the First and Second Observing Runs}",
    eprint = "1811.12907",
    archivePrefix = "arXiv",
    primaryClass = "astro-ph.HE",
    reportNumber = "LIGO-P1800307",
    doi = "10.1103/PhysRevX.9.031040",
    journal = "Phys. Rev. X",
    volume = "9",
    number = "3",
    pages = "031040",
    year = "2019"
}

@article{Landry:2021hvl,
    author = "Landry, Philippe and Read, Jocelyn S.",
    title = "{The Mass Distribution of Neutron Stars in Gravitational-wave Binaries}",
    eprint = "2107.04559",
    archivePrefix = "arXiv",
    primaryClass = "astro-ph.HE",
    doi = "10.3847/2041-8213/ac2f3e",
    journal = "Astrophys. J. Lett.",
    volume = "921",
    number = "2",
    pages = "L25",
    year = "2021"
}

@article{Martinez:2015mya,
    author = "Martinez, J. G. and Stovall, K. and Freire, P. C. C. and Deneva, J. S. and Jenet, F. A. and McLaughlin, M. A. and Bagchi, M. and Bates, S. D. and Ridolfi, A.",
    title = "{Pulsar J0453+1559: A Double Neutron Star System with a Large Mass Asymmetry}",
    eprint = "1509.08805",
    archivePrefix = "arXiv",
    primaryClass = "astro-ph.HE",
    doi = "10.1088/0004-637X/812/2/143",
    journal = "Astrophys. J.",
    volume = "812",
    number = "2",
    pages = "143",
    year = "2015"
}

@article{Kramer:2006nb,
    author = "Kramer, M. and others",
    title = "{Tests of general relativity from timing the double pulsar}",
    eprint = "astro-ph/0609417",
    archivePrefix = "arXiv",
    doi = "10.1126/science.1132305",
    journal = "Science",
    volume = "314",
    pages = "97--102",
    year = "2006"
}

@article{Fonseca:2014qla,
    author = "Fonseca, Emmanuel and Stairs, Ingrid H. and Thorsett, Stephen E.",
    title = "{A Comprehensive Study of Relativistic Gravity using PSR B1534+12}",
    eprint = "1402.4836",
    archivePrefix = "arXiv",
    primaryClass = "astro-ph.HE",
    doi = "10.1088/0004-637X/787/1/82",
    journal = "Astrophys. J.",
    volume = "787",
    pages = "82",
    year = "2014"
}

@article{Ferdman:2014rna,
    author = "Ferdman, Robert D. and others",
    title = "{PSR J1756{\ensuremath{-}}2251: a pulsar with a low-mass neutron star companion}",
    eprint = "1406.5507",
    archivePrefix = "arXiv",
    primaryClass = "astro-ph.SR",
    doi = "10.1093/mnras/stu1223",
    journal = "Mon. Not. Roy. Astron. Soc.",
    volume = "443",
    number = "3",
    pages = "2183--2196",
    year = "2014"
}

@article{vanLeeuwen:2014sca,
    author = "van Leeuwen, Joeri and others",
    title = "{The Binary Companion of Young, Relativistic Pulsar J1906+0746}",
    eprint = "1411.1518",
    archivePrefix = "arXiv",
    primaryClass = "astro-ph.SR",
    doi = "10.1088/0004-637X/798/2/118",
    journal = "Astrophys. J.",
    volume = "798",
    number = "2",
    pages = "118",
    year = "2015"
}

@article{Weisberg:2010zz,
    author = "Weisberg, J. M. and Nice, D. J. and Taylor, J. H.",
    title = "{Timing Measurements of the Relativistic Binary Pulsar PSR B1913+16}",
    eprint = "1011.0718",
    archivePrefix = "arXiv",
    primaryClass = "astro-ph.GA",
    doi = "10.1088/0004-637X/722/2/1030",
    journal = "Astrophys. J.",
    volume = "722",
    pages = "1030--1034",
    year = "2010"
}

@article{Jacoby:2006dy,
    author = "Jacoby, Bryan A. and Cameron, P. B. and Jenet, F. A. and Anderson, S. B. and Murty, R. N. and Kulkarni, S. R.",
    title = "{Measurement of Orbital Decay in the Double Neutron Star Binary PSR B2127+11C}",
    eprint = "astro-ph/0605375",
    archivePrefix = "arXiv",
    doi = "10.1086/505742",
    journal = "Astrophys. J. Lett.",
    volume = "644",
    pages = "L113--L116",
    year = "2006"
}

@article{Cameron:2017ody,
    author = "Cameron, A. D. and others",
    title = "{The High Time Resolution Universe Pulsar Survey {\textendash} XIII. PSR J1757{\ensuremath{-}}1854, the most accelerated binary pulsar}",
    eprint = "1711.07697",
    archivePrefix = "arXiv",
    primaryClass = "astro-ph.HE",
    doi = "10.1093/mnrasl/sly003",
    journal = "Mon. Not. Roy. Astron. Soc.",
    volume = "475",
    number = "1",
    pages = "L57--L61",
    year = "2018"
}

@article{Lynch:2018zxo,
    author = "Lynch, Ryan S. and others",
    title = "{The Green Bank North Celestial Cap Pulsar Survey III: 45 New Pulsar Timing Solutions}",
    eprint = "1805.04951",
    archivePrefix = "arXiv",
    primaryClass = "astro-ph.HE",
    doi = "10.3847/1538-4357/aabf8a",
    journal = "Astrophys. J.",
    volume = "859",
    number = "2",
    pages = "93",
    year = "2018"
}

@article{Ferdman:2020huz,
    author = "Ferdman, R. D. and others",
    title = "{Asymmetric mass ratios for bright double neutron-star mergers}",
    eprint = "2007.04175",
    archivePrefix = "arXiv",
    primaryClass = "astro-ph.HE",
    doi = "10.1038/s41586-020-2439-x",
    journal = "Nature",
    volume = "583",
    number = "7815",
    pages = "211--214",
    year = "2020"
}

@article{Bernadich:2023uru,
    author = "Bernadich, I, M. Colom and others",
    title = "{The MPIfR-MeerKAT Galactic Plane Survey - II. The eccentric double neutron star system PSR J1208{\ensuremath{-}}5936 and a neutron star merger rate update}",
    eprint = "2308.16802",
    archivePrefix = "arXiv",
    primaryClass = "astro-ph.HE",
    doi = "10.1051/0004-6361/202346953",
    journal = "Astron. Astrophys.",
    volume = "678",
    pages = "A187",
    year = "2023"
}

@article{Tan:2024zvy,
    author = "Tan, Chia Min and Fonseca, Emmanuel and Crowter, Kathryn and Dong, Fengqiu Adam and Kaspi, Victoria M. and Masui, Kiyoshi W. and McKee, James W. and Meyers, Bradley W. and Ransom, Scott M. and Stairs, Ingrid H.",
    title = "{High-cadence Timing of Binary Pulsars with CHIME}",
    eprint = "2402.08188",
    archivePrefix = "arXiv",
    primaryClass = "astro-ph.HE",
    doi = "10.3847/1538-4357/ad28b2",
    journal = "Astrophys. J.",
    volume = "966",
    number = "1",
    pages = "26",
    year = "2024"
}

@article{Haniewicz:2020jro,
    author = "Haniewicz, Henryk T. and Ferdman, Robert D. and Freire, Paulo C. C. and Champion, David J. and Bunting, Kaine A. and Lorimer, Duncan R. and McLaughlin, Maura A.",
    title = "{Precise mass measurements for the double neutron star system J1829+2456}",
    eprint = "2007.07565",
    archivePrefix = "arXiv",
    primaryClass = "astro-ph.SR",
    doi = "10.1093/mnras/staa3466",
    journal = "Mon. Not. Roy. Astron. Soc.",
    volume = "500",
    number = "4",
    pages = "4620--4627",
    year = "2020"
}

@article{Corongiu:2006rd,
    author = "Corongiu, Alessandro and Kramer, M. and Stappers, B. W. and Lyne, A. G. and Jessner, A. and Possenti, A. and D'Amico, N. and Loehmer, O.",
    title = "{The binary pulsar PSR J1811-1736: Evidence of a low amplitude supernova kick}",
    eprint = "astro-ph/0611436",
    archivePrefix = "arXiv",
    doi = "10.1051/0004-6361:20054385",
    journal = "Astron. Astrophys.",
    volume = "462",
    pages = "703",
    year = "2007"
}

@article{Swiggum:2015yra,
    author = "Swiggum, J. K. and others",
    title = "{PSR J1930-1852: a pulsar in the widest known orbit around another neutron star}",
    eprint = "1503.06276",
    archivePrefix = "arXiv",
    primaryClass = "astro-ph.HE",
    doi = "10.1088/0004-637X/805/2/156",
    journal = "Astrophys. J.",
    volume = "805",
    number = "2",
    pages = "156",
    year = "2015"
}

@article{Stovall:2018ouw,
    author = "Stovall, K. and others",
    title = "{PALFA Discovery of a Highly Relativistic Double Neutron Star Binary}",
    eprint = "1802.01707",
    archivePrefix = "arXiv",
    primaryClass = "astro-ph.HE",
    doi = "10.3847/2041-8213/aaad06",
    journal = "Astrophys. J. Lett.",
    volume = "854",
    number = "2",
    pages = "L22",
    year = "2018"
}

@article{Su:2024gak,
    author = "Su, W. Q. and others",
    title = "{The FAST Galactic Plane Pulsar Snapshot Survey {\textendash} V. PSR J1901+0658 in a double neutron star system}",
    eprint = "2403.11635",
    archivePrefix = "arXiv",
    primaryClass = "astro-ph.HE",
    doi = "10.1093/mnras/stae888",
    journal = "Mon. Not. Roy. Astron. Soc.",
    volume = "530",
    number = "2",
    pages = "1506--1511",
    year = "2024"
}

@article{Zhao:2024twu,
    author = "Zhao, D. and others",
    title = "{A Relativistic Double Neutron Star Binary PSR J1846-0513}",
    doi = "10.3847/2041-8213/ad2fb3",
    journal = "Astrophys. J. Lett.",
    volume = "964",
    number = "1",
    pages = "L7",
    year = "2024"
}

@article{Martinez:2017jbp,
    author = "Martinez, J. G. and Stovall, K. and Freire, P. C. C. and Deneva, J. S. and Tauris, T. M. and Ridolfi, A. and Wex, N. and Jenet, F. A. and McLaughlin, M. A. and Bagchi, M.",
    title = "{Pulsar J1411+2551: A Low-mass Double Neutron Star System}",
    eprint = "1711.09804",
    archivePrefix = "arXiv",
    primaryClass = "astro-ph.HE",
    doi = "10.3847/2041-8213/aa9d87",
    journal = "Astrophys. J. Lett.",
    volume = "851",
    number = "2",
    pages = "L29",
    year = "2017"
}

@misc{Essick2025a,
  author       = {R. Essick and et al.},
  title        = {gw-distributions},
  howpublished = {\url{https://git.ligo.org/reed.essick/gw-distributions}},
  year         = {2025},
  note         = {GitLab repository}
}

@misc{Essick2025b,
  author       = {R. Essick},
  title        = {monte-carlo-vt},
  howpublished = {\url{https://git.ligo.org/reed.essick/monte-carlo-vt}},
  year         = {2025},
  note         = {GitLab repository}
}

@article{Khan:2016,
	doi = "10.1103/physrevd.93.044007",
	year = "2016",
	month = "feb",
	publisher = "American Physical Society ({APS})",
	volume = "93",
	number = "4",
	pages = "044007",
	numpages = "27",
	author = {Khan, Sebastian and Husa, Sascha and Hannam, Mark and Ohme, Frank and P\"urrer, Michael and Forteza, Xisco Jim\'enez and Boh\'e, Alejandro},
	title = "{Frequency-domain gravitational waves from nonprecessing black-hole binaries. {II}. A phenomenological model for the advanced detector era}",
	journal = "Physical Review D"
}

@article{Husa:2016,
	doi = "10.1103/physrevd.93.044006",
	year = "2016",
	month = "feb",
	publisher = "American Physical Society ({APS})",
	volume = "93",
	number = "4",
	pages = "044006",
	numpages = "19",
	author = {Husa, Sascha and Khan, Sebastian and Hannam, Mark and P\"urrer, Michael and Ohme, Frank and Forteza, Xisco Jim\'enez and Boh\'e, Alejandro},
	title = "{Frequency-domain gravitational waves from nonprecessing black-hole binaries. I. New numerical waveforms and anatomy of the signal}",
	journal = "Physical Review D"
}

@article{Zhang:2019bhn,
    author = "Zhang, Jianwei and Yang, Yiyan and Zhang, Chengmin and Yang, Wuming and Li, Di and Bi, Shaolan and Zhang, Xianfei",
    title = "{The mass distribution of Galactic double neutron stars: constraints on the gravitational-wave sources like GW170817}",
    doi = "10.1093/mnras/stz2020",
    journal = "Mon. Not. Roy. Astron. Soc.",
    volume = "488",
    number = "4",
    pages = "5020--5028",
    year = "2019"
}

@article{Cannon:2012zt,
    author = "Cannon, Kipp and Hanna, Chad and Keppel, Drew",
    title = "{Method to estimate the significance of coincident gravitational-wave observations from compact binary coalescence}",
    eprint = "1209.0718",
    archivePrefix = "arXiv",
    primaryClass = "gr-qc",
    doi = "10.1103/PhysRevD.88.024025",
    journal = "Phys. Rev. D",
    volume = "88",
    number = "2",
    pages = "024025",
    year = "2013"
}

@phdthesis{Fong2018,
  author       = {Fong, H. K. Y.},
  title        = {From simulations to signals: Analyzing gravitational waves from compact binary coalescences},
  school       = {University of Toronto},
  year         = {2018},
  type         = {PhD thesis},
  url          = {http://hdl.handle.net/1807/91831},
}

@article{Owen:1995tm,
    author = "Owen, Benjamin J.",
    title = "{Search templates for gravitational waves from inspiraling binaries: Choice of template spacing}",
    eprint = "gr-qc/9511032",
    archivePrefix = "arXiv",
    doi = "10.1103/PhysRevD.53.6749",
    journal = "Phys. Rev. D",
    volume = "53",
    pages = "6749--6761",
    year = "1996"
}

@article{gwtc4,
    author = "Abac, A. G. and others",
    collaboration = "LIGO Scientific, VIRGO, KAGRA",
    title = "{GWTC-4.0: Updating the Gravitational-Wave Transient Catalog with Observations from the First Part of the Fourth LIGO-Virgo-KAGRA Observing Run}",
    eprint = "2508.18082",
    archivePrefix = "arXiv",
    primaryClass = "gr-qc",
    reportNumber = "LIGO-P2400386",
    doi = "10.3847/2041-8213/ae2c74",
    journal = "Astrophys. J. Lett.",
    volume = "1004",
    number = "2",
    pages = "L22",
    year = "2026"
}

@article{LIGOScientific:2026wfs,
    author = "Abac, A. G. and others",
    collaboration = "LIGO Scientific, VIRGO, KAGRA",
    title = "{GWTC-5.0: Observations from the Second Part of the Fourth LIGO-Virgo-KAGRA Observing Run and Updates to the Gravitational-Wave Transient Catalog}",
    eprint = "2605.27225",
    archivePrefix = "arXiv",
    primaryClass = "gr-qc",
    reportNumber = "LIGO-P2600152",
    month = "5",
    year = "2026"
}

@article{LIGOScientific:2017adf,
    author = "Abbott, B. P. and others",
    collaboration = "LIGO Scientific, Virgo, 1M2H, Dark Energy Camera GW-E, DES, DLT40, Las Cumbres Observatory, VINROUGE, MASTER",
    title = "{A gravitational-wave standard siren measurement of the Hubble constant}",
    eprint = "1710.05835",
    archivePrefix = "arXiv",
    primaryClass = "astro-ph.CO",
    reportNumber = "LIGO-P1700296, FERMILAB-PUB-17-472-A-AE",
    doi = "10.1038/nature24471",
    journal = "Nature",
    volume = "551",
    number = "7678",
    pages = "85--88",
    year = "2017"
}

@misc{LIGOScientific:2026ifv,
    author = "Abac, A. G. and others",
    collaboration = "LIGO Scientific, VIRGO, KAGRA",
    title = "{GWTC-5.0: Methods for Identifying and Characterizing Gravitational-wave Transients}",
    eprint = "2605.27224",
    archivePrefix = "arXiv",
    primaryClass = "gr-qc",
    reportNumber = "LIGO-P2600166",
    month = "5",
    year = "2026"
}

@article{gwtc4-methods,
    author = "Abac, A. G. and others",
    collaboration = "LIGO Scientific, VIRGO, KAGRA",
    title = "{GWTC-4.0: Methods for Identifying and Characterizing Gravitational-wave Transients}",
    journal = "arXiv e-prints",
    eprint = "2508.18081",
    archivePrefix = "arXiv",
    primaryClass = "gr-qc",
    reportNumber = "LIGO-P2400300",
    month = "8",
    year = "2025"
}

@misc{gwtc4-intro,
    author = "Abac, A. G. and others",
    collaboration = "LIGO Scientific, VIRGO, KAGRA",
    title = "{GWTC-4.0: An Introduction to Version 4.0 of the Gravitational-Wave Transient Catalog}",
    journal = "arXiv e-prints",
    eprint = "2508.18080",
    archivePrefix = "arXiv",
    primaryClass = "gr-qc",
    reportNumber = "LIGO-P2400293",
    month = "8",
    year = "2025"
}

@article{gwtc4-pop,
    author = "Abac, A. G. and others",
    collaboration = "LIGO Scientific, VIRGO, KAGRA",
    title = "{GWTC-4.0: Population Properties of Merging Compact Binaries}",
    journal = "arXiv e-prints",
    eprint = "2508.18083",
    archivePrefix = "arXiv",
    primaryClass = "astro-ph.HE",
    reportNumber = "LIGO-P2400004",
    month = "8",
    year = "2025"
}

@article{bilby_paper,
    author = "Ashton, Gregory and others",
    title = "{BILBY: A user-friendly Bayesian inference library for gravitational-wave astronomy}",
    eprint = "1811.02042",
    archivePrefix = "arXiv",
    primaryClass = "astro-ph.IM",
    doi = "10.3847/1538-4365/ab06fc",
    journal = "Astrophys. J. Suppl.",
    volume = "241",
    number = "2",
    pages = "27",
    year = "2019"
}

@article{bilby_pipe_paper,
    author = "Romero-Shaw, I. M. and others",
    title = "{Bayesian inference for compact binary coalescences with bilby: validation and application to the first LIGO\textendash{}Virgo gravitational-wave transient catalogue}",
    eprint = "2006.00714",
    archivePrefix = "arXiv",
    primaryClass = "astro-ph.IM",
    doi = "10.1093/mnras/staa2850",
    journal = "Mon. Not. Roy. Astron. Soc.",
    volume = "499",
    number = "3",
    pages = "3295--3319",
    year = "2020"
}

@article{Dietrich:2019kaq,
    author = "Dietrich, Tim and Samajdar, Anuradha and Khan, Sebastian and Johnson-McDaniel, Nathan K. and Dudi, Reetika and Tichy, Wolfgang",
    title = "{Improving the NRTidal model for binary neutron star systems}",
    eprint = "1905.06011",
    archivePrefix = "arXiv",
    primaryClass = "gr-qc",
    doi = "10.1103/PhysRevD.100.044003",
    journal = "Phys. Rev. D",
    volume = "100",
    number = "4",
    pages = "044003",
    year = "2019"
}

@article{Colleoni:2023ple,
    author = "Colleoni, Marta and Ramis Vidal, Felip A. and Johnson-McDaniel, Nathan K. and Dietrich, Tim and Haney, Maria and Pratten, Geraint",
    title = "{New gravitational waveform model for precessing binary neutron stars with double-spin effects}",
    eprint = "2311.15978",
    archivePrefix = "arXiv",
    primaryClass = "gr-qc",
    doi = "10.1103/PhysRevD.111.064025",
    journal = "Phys. Rev. D",
    volume = "111",
    number = "6",
    pages = "064025",
    year = "2025"
}

@article{Astropy:2022ucr,
    author = "Price-Whelan, Adrian M. and others",
    collaboration = "Astropy",
    title = "{The Astropy Project: Sustaining and Growing a Community-oriented Open-source Project and the Latest Major Release (v5.0) of the Core Package*}",
    eprint = "2206.14220",
    archivePrefix = "arXiv",
    primaryClass = "astro-ph.IM",
    doi = "10.3847/1538-4357/ac7c74",
    journal = "Astrophys. J.",
    volume = "935",
    number = "2",
    pages = "167",
    year = "2022"
}

@article{Planck:2015fie,
    author = "Ade, P. A. R. and others",
    collaboration = "Planck",
    title = "{Planck 2015 results. XIII. Cosmological parameters}",
    eprint = "1502.01589",
    archivePrefix = "arXiv",
    primaryClass = "astro-ph.CO",
    doi = "10.1051/0004-6361/201525830",
    journal = "Astron. Astrophys.",
    volume = "594",
    pages = "A13",
    year = "2016"
}

@article{Chatziioannou:2018vzf,
    author = "Chatziioannou, Katerina and Haster, Carl-Johan and Zimmerman, Aaron",
    title = "{Measuring the neutron star tidal deformability with equation-of-state-independent relations and gravitational waves}",
    eprint = "1804.03221",
    archivePrefix = "arXiv",
    primaryClass = "gr-qc",
    doi = "10.1103/PhysRevD.97.104036",
    journal = "Phys. Rev. D",
    volume = "97",
    number = "10",
    pages = "104036",
    year = "2018"
}

@article{Gupta:2023jrn,
    author = "Gupta, Toral and Cornish, Neil J.",
    title = "{Bayesian power spectral estimation of gravitational wave detector noise revisited}",
    eprint = "2312.11808",
    archivePrefix = "arXiv",
    primaryClass = "gr-qc",
    doi = "10.1103/PhysRevD.109.064040",
    journal = "Phys. Rev. D",
    volume = "109",
    number = "6",
    pages = "064040",
    year = "2024"
}

@article{Hannam:2013oca,
    author = {Hannam, Mark and Schmidt, Patricia and Boh{\'e}, Alejandro and Haegel, Le{\"\i}la and Husa, Sascha and Ohme, Frank and Pratten, Geraint and P{\"u}rrer, Michael},
    title = "{Simple Model of Complete Precessing Black-Hole-Binary Gravitational Waveforms}",
    eprint = "1308.3271",
    archivePrefix = "arXiv",
    primaryClass = "gr-qc",
    doi = "10.1103/PhysRevLett.113.151101",
    journal = "Phys. Rev. Lett.",
    volume = "113",
    number = "15",
    pages = "151101",
    year = "2014"
}

@article{Khan:2015jqa,
    author = {Khan, Sebastian and Husa, Sascha and Hannam, Mark and Ohme, Frank and P{\"u}rrer, Michael and Jim{\'e}nez Forteza, Xisco and Boh{\'e}, Alejandro},
    title = "{Frequency-domain gravitational waves from nonprecessing black-hole binaries. II. A phenomenological model for the advanced detector era}",
    eprint = "1508.07253",
    archivePrefix = "arXiv",
    primaryClass = "gr-qc",
    doi = "10.1103/PhysRevD.93.044007",
    journal = "Phys. Rev. D",
    volume = "93",
    number = "4",
    pages = "044007",
    year = "2016"
}

@article{Littenberg:2014oda,
    author = "Littenberg, Tyson B. and Cornish, Neil J.",
    title = "{Bayesian inference for spectral estimation of gravitational wave detector noise}",
    eprint = "1410.3852",
    archivePrefix = "arXiv",
    primaryClass = "gr-qc",
    doi = "10.1103/PhysRevD.91.084034",
    journal = "Phys. Rev. D",
    volume = "91",
    number = "8",
    pages = "084034",
    year = "2015"
}

@article{Morisaki:2023kuq,
    author = "Morisaki, Soichiro and Smith, Rory and Tsukada, Leo and Sachdev, Surabhi and Stevenson, Simon and Talbot, Colm and Zimmerman, Aaron",
    title = "{Rapid localization and inference on compact binary coalescences with the Advanced LIGO-Virgo-KAGRA gravitational-wave detector network}",
    eprint = "2307.13380",
    archivePrefix = "arXiv",
    primaryClass = "gr-qc",
    doi = "10.1103/PhysRevD.108.123040",
    journal = "Phys. Rev. D",
    volume = "108",
    number = "12",
    pages = "123040",
    year = "2023"
}

@article{Smith:2016qas,
    author = {Smith, Rory and Field, Scott E. and Blackburn, Kent and Haster, Carl-Johan and P{\"u}rrer, Michael and Raymond, Vivien and Schmidt, Patricia},
    title = "{Fast and accurate inference on gravitational waves from precessing compact binaries}",
    eprint = "1604.08253",
    archivePrefix = "arXiv",
    primaryClass = "gr-qc",
    reportNumber = "LIGO-DOCUMENT-NUMBER-P1600096, LIGO-P1600096",
    doi = "10.1103/PhysRevD.94.044031",
    journal = "Phys. Rev. D",
    volume = "94",
    number = "4",
    pages = "044031",
    year = "2016"
}

@article{Yagi:2015pkc,
    author = "Yagi, Kent and Yunes, Nicolas",
    title = "{Binary Love Relations}",
    eprint = "1512.02639",
    archivePrefix = "arXiv",
    primaryClass = "gr-qc",
    doi = "10.1088/0264-9381/33/13/13LT01",
    journal = "Class. Quant. Grav.",
    volume = "33",
    number = "13",
    pages = "13LT01",
    year = "2016"
}

@article{Pian:2017gtc,
    author = "Pian, E. and others",
    title = "{Spectroscopic identification of r-process nucleosynthesis in a double neutron star merger}",
    eprint = "1710.05858",
    archivePrefix = "arXiv",
    primaryClass = "astro-ph.HE",
    doi = "10.1038/nature24298",
    journal = "Nature",
    volume = "551",
    pages = "67--70",
    year = "2017"
}

@article{Drout:2017ijr,
    author = "Drout, M. R. and others",
    title = "{Light Curves of the Neutron Star Merger GW170817/SSS17a: Implications for R-Process Nucleosynthesis}",
    eprint = "1710.05443",
    archivePrefix = "arXiv",
    primaryClass = "astro-ph.HE",
    doi = "10.1126/science.aaq0049",
    journal = "Science",
    volume = "358",
    pages = "1570--1574",
    year = "2017"
}

@article{Capote:2024rmo,
    author = "Capote, E. and others",
    title = "{Advanced LIGO detector performance in the fourth observing run}",
    eprint = "2411.14607",
    archivePrefix = "arXiv",
    primaryClass = "gr-qc",
    reportNumber = "LIGO-P2400256",
    doi = "10.1103/PhysRevD.111.062002",
    journal = "Phys. Rev. D",
    volume = "111",
    number = "6",
    pages = "062002",
    year = "2025"
}

@article{LIGO:2024kkz,
    author = "Soni, S. and others",
    collaboration = "LIGO",
    title = "{LIGO Detector Characterization in the first half of the fourth Observing run}",
    eprint = "2409.02831",
    archivePrefix = "arXiv",
    primaryClass = "astro-ph.IM",
    doi = "10.1088/1361-6382/adc4b6",
    journal = "Class. Quant. Grav.",
    volume = "42",
    number = "8",
    pages = "085016",
    year = "2025"
}

@article{Alsing:2017bbc,
    author = "Alsing, Justin and Silva, Hector O. and Berti, Emanuele",
    title = "{Evidence for a maximum mass cut-off in the neutron star mass distribution and constraints on the equation of state}",
    eprint = "1709.07889",
    archivePrefix = "arXiv",
    primaryClass = "astro-ph.HE",
    doi = "10.1093/mnras/sty1065",
    journal = "Mon. Not. Roy. Astron. Soc.",
    volume = "478",
    number = "1",
    pages = "1377--1391",
    year = "2018"
}

@article{Nelemans:2003xp,
    author = "Nelemans, G.",
    editor = "Centrella, J. M.",
    title = "{Galactic binaries as sources of gravitational waves}",
    eprint = "astro-ph/0310800",
    archivePrefix = "arXiv",
    doi = "10.1063/1.1629441",
    journal = "AIP Conf. Proc.",
    volume = "686",
    number = "1",
    pages = "263--272",
    year = "2003"
}

@article{Ye:2021klk,
    author = "Ye, Christine and Fishbach, Maya",
    title = "{Cosmology with standard sirens at cosmic noon}",
    eprint = "2103.14038",
    archivePrefix = "arXiv",
    primaryClass = "astro-ph.CO",
    doi = "10.1103/PhysRevD.104.043507",
    journal = "Phys. Rev. D",
    volume = "104",
    number = "4",
    pages = "043507",
    year = "2021"
}

@article{LIGOScientific:2017ync,
    author = "Abbott, B. P. and others",
    collaboration = "LIGO Scientific, Virgo, Fermi GBM, INTEGRAL, IceCube, AstroSat Cadmium Zinc Telluride Imager Team, IPN, Insight-Hxmt, ANTARES, Swift, AGILE Team, 1M2H Team, Dark Energy Camera GW-EM, DES, DLT40, GRAWITA, Fermi-LAT, ATCA, ASKAP, Las Cumbres Observatory Group, OzGrav, DWF (Deeper Wider Faster Program), AST3, CAASTRO, VINROUGE, MASTER, J-GEM, GROWTH, JAGWAR, CaltechNRAO, TTU-NRAO, NuSTAR, Pan-STARRS, MAXI Team, TZAC Consortium, KU, Nordic Optical Telescope, ePESSTO, GROND, Texas Tech University, SALT Group, TOROS, BOOTES, MWA, CALET, IKI-GW Follow-up, H.E.S.S., LOFAR, LWA, HAWC, Pierre Auger, ALMA, Euro VLBI Team, Pi of Sky, Chandra Team at McGill University, DFN, ATLAS Telescopes, High Time Resolution Universe Survey, RIMAS, RATIR, SKA South Africa/MeerKAT",
    title = "{Multi-messenger Observations of a Binary Neutron Star Merger}",
    eprint = "1710.05833",
    archivePrefix = "arXiv",
    primaryClass = "astro-ph.HE",
    reportNumber = "LIGO-P1700294, VIR-0802A-17, FERMILAB-PUB-17-478-A-AE-CD",
    doi = "10.3847/2041-8213/aa91c9",
    journal = "Astrophys. J. Lett.",
    volume = "848",
    number = "2",
    pages = "L12",
    year = "2017"
}

@article{Essick:2025zed,
    author = "Essick, Reed and others",
    title = "{Compact binary coalescence sensitivity estimates with injection campaigns during the LIGO-Virgo-KAGRA Collaborations{\textquoteright} fourth observing run}",
    eprint = "2508.10638",
    archivePrefix = "arXiv",
    primaryClass = "gr-qc",
    doi = "10.1103/44x3-hv3y",
    journal = "Phys. Rev. D",
    volume = "112",
    number = "10",
    pages = "102001",
    year = "2025"
}

@article{Farr:2013yna,
    author = "Farr, Will M. and Gair, Jonathan R. and Mandel, Ilya and Cutler, Curt",
    title = "{Counting And Confusion: Bayesian Rate Estimation With Multiple Populations}",
    eprint = "1302.5341",
    archivePrefix = "arXiv",
    primaryClass = "astro-ph.IM",
    doi = "10.1103/PhysRevD.91.023005",
    journal = "Phys. Rev. D",
    volume = "91",
    number = "2",
    pages = "023005",
    year = "2015"
}

@article{LIGOScientific:2020kqk,
    author = "Abbott, R. and others",
    collaboration = "LIGO Scientific, Virgo",
    title = "{Population Properties of Compact Objects from the Second LIGO-Virgo Gravitational-Wave Transient Catalog}",
    eprint = "2010.14533",
    archivePrefix = "arXiv",
    primaryClass = "astro-ph.HE",
    reportNumber = "LIGO-P2000077",
    doi = "10.3847/2041-8213/abe949",
    journal = "Astrophys. J. Lett.",
    volume = "913",
    number = "1",
    pages = "L7",
    year = "2021"
}

@dataset{LIGO:2025gwtc4vt,
  author       = {{LVK Collaboration}},
  title        = {{GWTC-4.0 Cumulative Search Sensitivity Estimates}},
  year         = {2025},
  publisher    = {Zenodo},
  doi          = {10.5281/zenodo.16740128}
}

@article{Khadkikar:2025awf,
    author = "Khadkikar, Sanika and Singh, Divya",
    title = "{Case study of GW190425 for classifying binary neutron star versus binary black hole mergers and constraining asymmetric dark matter with gravitational wave detectors}",
    eprint = "2507.07895",
    archivePrefix = "arXiv",
    primaryClass = "astro-ph.HE",
    reportNumber = "LIGO DCC Number: P2500417-v2",
    doi = "10.1103/tzz8-wvrf",
    journal = "Phys. Rev. D",
    volume = "113",
    number = "2",
    pages = "023029",
    year = "2026"
}

@article{Cannon:2015gha,
    author = "Cannon, Kipp and Hanna, Chad and Peoples, Jacob",
    title = "{Likelihood-Ratio Ranking Statistic for Compact Binary Coalescence Candidates with Rate Estimation}",
    eprint = "1504.04632",
    archivePrefix = "arXiv",
    primaryClass = "astro-ph.IM",
    reportNumber = "LIGO-P1400175",
    month = "4",
    year = "2015"
}

@article{LIGOScientific:2025snk,
    author = "Abac, A. G. and others",
    collaboration = "LIGO Scientific, Virgo, KAGRA",
    title = "{Open Data from LIGO, Virgo, and KAGRA through the First Part of the Fourth Observing Run}",
    eprint = "2508.18079",
    archivePrefix = "arXiv",
    primaryClass = "gr-qc",
    reportNumber = "LIGO-P2500167",
    doi = "10.3847/1538-4357/ae211e",
    journal = "Astrophys. J.",
    volume = "1004",
    number = "2",
    pages = "232",
    year = "2026"
}

@article{Margalit:2017dij,
    author = "Margalit, Ben and Metzger, Brian D.",
    title = "{Constraining the Maximum Mass of Neutron Stars From Multi-Messenger Observations of GW170817}",
    eprint = "1710.05938",
    archivePrefix = "arXiv",
    primaryClass = "astro-ph.HE",
    doi = "10.3847/2041-8213/aa991c",
    journal = "Astrophys. J. Lett.",
    volume = "850",
    number = "2",
    pages = "L19",
    year = "2017"
}

@ARTICLE{Tauris2017,
       author = {{Tauris}, T. M. and {Kramer}, M. and {Freire}, P. C. C. and {Wex}, N. and {Janka}, H.-T. and {Langer}, N. and {Podsiadlowski}, Ph. and {Bozzo}, E. and {Chaty}, S. and {Kruckow}, M. U. and {van den Heuvel}, E. P. J. and {Antoniadis}, J. and {Breton}, R. P. and {Champion}, D. J.},
        title = "{Formation of Double Neutron Star Systems}",
      journal = {The Astrophysical Journal},
         year = 2017,
        month = sep,
       volume = {846},
       number = {2},
          eid = {170},
        pages = {170},
          doi = {10.3847/1538-4357/aa7e89},
       adsurl = {https://ui.adsabs.harvard.edu/abs/2017ApJ...846..170T}
}

@article{Chattaraj:2025rig,
    author = "Chattaraj, Abhishek and others",
    title = "{Forming Double Neutron Stars Using Detailed Binary Evolution Models with POSYDON: Comparison to the Galactic Systems}",
    eprint = "2508.00186",
    archivePrefix = "arXiv",
    primaryClass = "astro-ph.SR",
    doi = "10.3847/1538-4357/ae1f93",
    journal = "Astrophys. J.",
    volume = "997",
    number = "1",
    pages = "52",
    year = "2026"
}

@ARTICLE{Schwab2010,
       author = {{Schwab}, J. and {Podsiadlowski}, Ph. and {Rappaport}, S.},
        title = "{Further Evidence for the Bimodal Distribution of Neutron-star Masses}",
      journal = {The Astrophysical Journal},
         year = 2010,
        month = aug,
       volume = {719},
       number = {1},
        pages = {722-727},
          doi = {10.1088/0004-637X/719/1/722},
       adsurl = {https://ui.adsabs.harvard.edu/abs/2010ApJ...719..722S}
}

@article{Nair:2026byv,
    author = "Nair, Ashwathi and Stevenson, Simon",
    title = "{Formation of heavy double neutron stars II: the role of heavy first-born neutron stars and low metallicity}",
    eprint = "2608.17430",
    archivePrefix = "arXiv",
    primaryClass = "astro-ph.SR",
    month = "8",
    year = "2026"
}

@article{Nair:2025kqv,
    author = "Nair, Ashwathi and Stevenson, Simon",
    title = "{Formation of heavy double neutron stars {\textendash} I. Eddington-limited accretion for a 1.4~M{\ensuremath{\odot}} neutron star at solar metallicity}",
    eprint = "2508.15624",
    archivePrefix = "arXiv",
    primaryClass = "astro-ph.HE",
    doi = "10.1093/mnras/staf1397",
    journal = "Mon. Not. Roy. Astron. Soc.",
    volume = "543",
    number = "1",
    pages = "233--246",
    year = "2025"
}

\end{document}